\documentclass[sn-mathphys,Numbered]{sn-jnl}

\usepackage{graphicx}%
\usepackage{multirow}%
\usepackage{amsmath,amssymb,amsfonts}%
\usepackage{amsthm}%
\usepackage{mathrsfs}%
\usepackage[title]{appendix}%
\usepackage{xcolor}%
\usepackage{textcomp}%
\usepackage{manyfoot}%
\usepackage{booktabs}%
\usepackage{algorithm}%
\usepackage{algorithmicx}%
\usepackage{algpseudocode}%
\usepackage{listings}%
\usepackage{aas_macros}

\theoremstyle{thmstyleone}%
\theoremstyle{thmstyletwo}%

\theoremstyle{thmstylethree}%

\begin{document}
\newcommand{\referee}[1]{\color{black}#1}

\title[Citizen Science in European Research Infrastructures]{Citizen Science in European Research Infrastructures}


\author*[1]{\fnm{Stephen} \sur{Serjeant}}\email{stephen.serjeant@open.ac.uk}

\author[1]{\fnm{James} \sur{Pearson}}\email{jpearson42@outlook.com}

\author[1]{\fnm{Hugh} \sur{Dickinson}}\email{hugh.dickinson@open.ac.uk}

\author[1]{\fnm{Johanna} \sur{Jarvis}}\email{johanna@thejarvishouse.co.uk}

\affil*[1]{\orgdiv{School of Physical Sciences}, \orgname{The Open University}, \orgaddress{\street{Walton Hall}, \city{Milton Keynes}, \postcode{MK7 6AA}, \country{UK}}}




\abstract{Major European Union-funded research infrastructure and open science projects have traditionally included dissemination work, for mostly one-way communication of the research activities. Here we present and review our radical re-envisioning of this work, by directly engaging citizen science volunteers into the research. We summarise the citizen science in the Horizon-funded projects ASTERICS (Astronomy ESFRI and Research Infrastructure Clusters) and ESCAPE (European Science Cluster of Astronomy and Particle Physics ESFRI Research Infrastructures), engaging hundreds of thousands of volunteers in providing millions of data mining classifications. Not only does this have enormously more scientific and societal impact than conventional dissemination, but it facilitates the direct research involvement of what is often arguably the most neglected stakeholder group in Horizon projects, the science-inclined public. We conclude with recommendations and opportunities for deploying crowdsourced data mining in the physical sciences, noting that the primary goal is always the fundamental research question; if public engagement is the primary goal to optimise, then other, more targeted approaches may be more effective. 
}




\maketitle
\section{Introduction}\label{sec:intro}

There has been an explosion in the volume and complexity of scientific data. Many scientific disciplines have begun a new data-rich era of discovery, from genomics to astronomical imaging to social sciences working with the firehose of data from social networks. But the data avalanche is so fast, so large and so complex that it often presents serious challenges for computing. Machine learning / artificial intelligence can alleviate these challenges, but this is usually contingent on having sufficient labelled data sets to act as training sets for machine learning. Unsupervised machine learning technologies are often insufficient to the task. 

Humans are still often much better than machine learning at many classification tasks. This has led to a new way of doing science: crowdsourcing the data mining, with the help of citizen science volunteers. Indeed some open science data mining problems are at present intrinsically intractable even for machine learning, particularly where a subjective human assessment is an essential element of the problem at hand. Furthermore, even the most sophisticated machine learning / artificial intelligence (ML/AI) technologies are unable to respond to a classification problem with, in effect, “After seeing the data, I now believe this project is asking the wrong question, because the data appear to be showing us something unexpected”. For example, Serjeant \cite{Serjeant2023} 
presented a hypothetical example of a facial recognition machine learning algorithm being presented with images of commuters. Such an algorithm would report the positions and possibly the identities of the faces, but if there were, for example, a circus clown working their way through the commuters, the algorithm would fail to identify the unusual occurrence. A human volunteer, however, would be much more likely to spot the outlier in the data. This capacity also makes human volunteers much better at classifying data where the subjects of interest are too rare, or the sample sizes too small, for machine learning. One of the best examples of this is from the Snapshot Serengeti citizen science project, which asked volunteers to identify wildlife from webcam images in a nature reserve \cite{Swanson2015}. 
This project identified over 100 thousand wildebeest, for which there are now machine learning algorithms that could replicate the effort, but the project also found 17 examples of the rare and elusive zorilla. 

{\referee There is no realistic prospect that machine learning will be able to entirely remove the need for human classifications, for multiple reasons. Firstly, machine learning is only as good as its training data, and where classification subjects are rare (as in the zorilla problem above), by definition the machine learning will fail. Secondly, despite the name ``artificial intelligence'', there is never any underlying intelligence of any sort. Large language models are now infamous for producing lucid text that is both confident and wrong, which is in turn due to there being no underlying knowledge model. Thirdly the internal operations of machine learning are often opaque, to the extent that entire computer science research themes are aimed at unpicking these internal operations. Pernicious biases have been discovered in many machine learning classification outputs, which can sometimes have serious consequences (e.g. facial recognition trained on biased training sets), while human volunteers are much more able to step outside the question posed (see the clown example above).}

The science-inclined public is also by far the largest, but often most overlooked, set of stakeholders in open science. A central vision of the European Open Science Cloud (EOSC) is to make scientific data FAIR, that is Findable, Accessible, Interoperable and Reusable. Implicit in this vision is that FAIR data should also be useful, but this is far from being guaranteed, especially given its inter- and multi-disciplinary remit. There are many examples of errors in FAIR data use by researchers outside their direct specialism. The farther from one’s subject specialism, the more curated one’s interaction with data needs to be. For example, Daylan et al. \cite{Daylan2016}
reanalysed public data from the Fermi gamma-ray telescope taken in the direction of the Galactic centre, where the dark matter particle density is predicted to be the highest in the Milky Way halo. The authors found a gamma-ray excess, which they interpreted as a signature of dark matter particle annihilation, which in turn would represent a major insight into the nature of dark matter particles. However, the instrument team re-interpreted this excess as an observational systematic 
\cite{Ackermann2017}.
Without taking any view on this technical debate, it is clear that the usefulness of FAIR data will always be limited by the supporting contextual information. This is sometimes glibly incorporated into expectations on metadata, but it is clear from this example that this may even extend to user training. The further away that you are working from your subject specialism, the more curated your interaction must be with data. Perhaps the most extreme example of this is the science-inclined public, who need the most support in their interaction with FAIR data.

With a suitably carefully curated interaction with FAIR data, members of the public can make a genuine and valuable participation in scientific discovery. Furthermore, there is a huge public appetite for taking part. Therefore, our approach has been to involve the science-inclined public with the research infrastructures (and their precursors/pathfinders) and open science projects with data mining and data collection citizen science projects. In this paper, we will summarise the efforts taken towards these goals in several large European Union funde projects: ASTERICS (Astronomy ESFRI and Research Infrastructure Clusters), 
ESCAPE (European Science Cluster of Astronomy and Particle Physics ESFRI
Research Infrastructures), and the EOSC-Future project that aimed to develop services for the European Open Science Cloud. 

\section{Methodology: existing infrastructure}\label{sec:method}
There are a number of citizen science platforms suitable in principle for integration with research infrastructure and/or EOSC services. In this section we will review a selection of the current marketplace of these platforms, and the use cases for which each platform is often deployed and perhaps best suited. We focus on data mining and data collection tasks where the majority of the end users are the science-inclined public, despite projects being managed by core teams of subject specialist academics. For the sake of brevity, we do not discuss policy or training portals such as EU-Citizen.Science. 

The Zooniverse has a very wide range of data mining capabilities for researchers to tailor for their non-specialist volunteers. Volunteer workflows can include polling among options, free text entry, image annotation of features or regions of interest, and more. Volunteers may be given options of reversing the colour balances of videos, and may pan or scroll through images. Short movies can be included as subjects using e.g. animated gifs, and the platform also supports audio clips. This is enough scope to facilitate a wide range of scientific capabilities, including transcription of handwritten texts, identification of objects or features in images, classification of audio and/or visual data, and more. The reach of the Zooniverse platform is enormous with over 2 million volunteers. In the 2019-20 year alone\footnote{Source: \url{https://blog.zooniverse.org/2020/11/17/into-the-zooniverse-vol-ii-now-available/}}, Zooniverse hosted 65 new projects, with volunteers contributing over 85 million classifications. In that year alone the research teams published 35 papers. The platform grew from a desire to generalise the successes of the Galaxy Zoo project, and perhaps partly as a result of that heritage from astronomy and astrophysics, the Zooniverse has functionalities that are well-tested in ESCAPE science domains. The platform also supports project site translations of navigation interfaces in some languages (researchers running projects must still supply translations of project-specific text), making the platform suitable for the international contexts of EOSC. Therefore, the Zooniverse platform was ESCAPE’s obvious first choice for integration into EOSC. 

The Zooniverse is well suited to data mining problems, but a host of other platforms have specialised in the different requirements of data collection. For example, the nQuire site has capabilities for confidentially surveying volunteers themselves, and for uploading audio recordings,  images or movies. Volunteer surveys can include free text entry, numerical entry, checkboxes or radio buttons, as well as supporting data from external sensors such as accelerometers, gyroscopes, orientation indicators, ambient light sensors, magnetometers, audio volumes, and location trackers. 

Another platform specialising in data collection is CitSci.org, which now hosts 1249 projects, which in turn have contributed 1\,825\,928 data points. The platform offers volunteers the capability of uploading images and observations on a wide range of devices, plus some options for volunteers to visualise and mine the data. 

A more specialised free platform is Treezilla, which aims to map, measure and monitor trees across the UK and the Republic of Ireland, with the help of public volunteers as well as local authorities, businesses and other organisations. Another UK-based specialised biodiversity project enrolling non-specialists is the Big Butterfly Count. Both this and Treezilla are single bespoke solutions for individual projects, in contrast to nQuire or CitSci.org. SPOTTERON is a platform that specialises in environmental monitoring for a wide range of citizen science projects, but the platform is not free at the point of use for researchers. 

Efforts have also been made to integrate data collection functionality into the Zooniverse. Rather than duplicating the functionalities of other platforms, Zooniverse partnered with the CitSci.org platform mentioned above. As with many technical developments in citizen science, this was driven by a direct scientific need, in this case from the Mountain Goat Molt Project. This project was crowdsourcing image data of goats at various locations and times, but wished to also crowdsource aspects of the image analysis. With the benefit of funding from the US National Science Foundation, projects on CitSci.org and Zooniverse can now be linked, and data from CitSci.org can be uploaded in near-real time for analysis in Zooniverse\footnote{Source: \url{https://blog.citsci.org/2021/11/17/crowdsource-images-with-citsci-analyze-them-with-the-zooniverse/}}. 

There are other ways for non-specialist volunteers to participate in scientific discovery in open science, beyond data mining and data collection. One of the earliest examples of this is the SETI@Home project, in which volunteers downloaded a screensaver that mined the SETI data sets, so volunteers donate otherwise idle CPU time for the science objectives of the SETI institute. As the balance changed between crowdsourced computation and the costs of high performance computing, the focus has shifted to computationally-harder problems that involve user interaction. For example, the Foldit project gamifies the discovery of protein folding configurations. Protein folding problem is an NP-hard problem (non deterministic polynomial hard), so it is not expected that there can be a general polynomial-time algorithm for finding the minimum energy configurations, although insights may be possible in individual cases. This is one of the few citizen science projects that involve elements of explicit competition between volunteers. At the time of writing, machine learning advances in protein structure prediction such as AlphaFold have not yet removed the need for testing against Foldit structure solutions\footnote{See e.g.: \url{https://fold.it/forum/blog/alphafold-machine-learning-for-protein-structure-prediction}}, as well as against experimental data; meanwhile, Foldit has also been extending to other areas such as protein design.  

\section{Results}\label{sec:results}

\subsection{Overview}

The EU-funded projects included deliverables to create a number of citizen science project demonstrators. This section describes the projects created during the funded lifetime of the ESCAPE and ASTERICS, and the extended work performed with the benefit of funding from the EOSC-Future project. Going beyond the baseline requirements of the minimum viable product of these demonstrators, we have also used several of them as exemplars in tutorial notebooks for creating and managing citizen science data mining projects, and integrating real-time machine learning into the projects; these tools will be described later in section \ref{sec:tools}. 

\subsection{Muon Hunter and Muon Hunters 2.0 - Return of the rings}\label{sec:muonhunter} 


The VERITAS team had been particularly keen to launch a major new citizen science experiment in the run-up to the Cherenkov Telescope Array. 
The original Muon Hunter experiment was therefore based on data from the VERITAS telescopes, used to detect some of the highest-energy photons in the Universe. These gamma-ray photons are generated  in extreme astrophysical environments such as the expanding blast waves driven by supernova explosions, or from relativistic jets in active galaxies. Muons are not the target astrophysical signal, but rather are a prominent background contaminant when observing very-high-energy gamma rays on terrestrial facilities. Muons can be distinguished in principle from the gamma-ray signals, because the muons present a distinctive ring-like morphology on the detectors. These can be relatively easy for a human to discern or detect, but incomplete or truncated rings can appear as false-positive gamma-ray signals to automatic analysis algorithms. We therefore sought the help of the public to identify camera images that contain muon rings to generate a truth set for training machine learning algorithms to remove these false positive signals. This project was a runaway success. This original Muon Hunters project welcomed 6107 citizen scientists who made 2\,161\,338 classifications of 135\,000 objects. In its first five days it attracted 1.3 million classifications, making it one of the most successful projects to run on the Zooniverse platform.

This machine learning led directly to the follow-up project, Muon Hunters 2.0, which tested the automated classifications, and sought to build a virtuous circle between human and machine classifications. The project also changed the manner in which the data was presented to volunteers, showing images in a grid pattern in order to bring additional efficiency to the classifications. Results from these projects are summarised in Feng et al. \cite{Feng2017}, Bird et al. \cite{Bird2018}, Laraia et al. \cite{Laraia2019}, Bird et al. \cite{Bird2020} and Flanagan et al. \cite{Flanagan2022}.

\subsection{Classifying Variable Stars using SuperWASP data}\label{sec:superwaspvariablestars}

This project aims to classify the many types of periodic variable stars found in SuperWASP survey data using the Zooniverse citizen science platform.

The Wide Angle Search for Planets, SuperWASP, is a leading ground-based survey data set for transiting hot Jupiters (massive gaseous planets). The project accumulated 16 million images from 2800 nights, comprising 580 billion data points from 31 million unique objects outside the Galactic plane, over the coures of 2004-2013. 

A re-analysis of all of the light curves led to the identification of 0.8 million unique objects have been identified as having statistically significant but unexplained variabilities. The volunteer tasks were to classify these variabilities, presented as folded light curves, as resembling that of an EA or EB type eclipsing binary star, an EW type eclipsing binary star, a pulsating star or a generic rotational modulation. 

This project has been enormously successful. At the time of writing, it has attracted 14\,052 volunteers who have contributed 5\,192\,629 classifications, covering 1\,851\,043 classification subjects. More details of the project results can be found in Norton \cite{Norton2018} and Thiemann et al. \cite{Thiemann2021}. 

\subsection{CREDO – Involving the public in scientific research}\label{sec:credo}

The Cosmic Ray Extremely Distributed Observatory (CREDO) collaboration is an ongoing research project involving scientists and the public from around the world. Unlike most of the projects discussed in this paper, the CREDO project is primarily a data {\it collection} exercise. The ``high risk, high return'' goals include the determination of the nature of dark matter. The ASTERICS project partially supported the creation of the CREDO project in its early stages, and since that time the project has flourished. 

The primary objective is to search for cascading products of the decay of supermassive particles, such as dark matter. These particle cascades occur elsewhere in the solar system so will be observed as being distributed over a very large international geographical area. CREDO is utilising the millions of small detectors throughout the globe in smartphone cameras, operating in a continuous readout of dark frames. These are supplemented by dark frames from worldwide astronomical observatories, and 
analysed by the public. The project platform development and science goals are described in more detail elsewhere \cite{credo2019,credo2022,credo2022b}.


\subsection{Challenge The Machines}\label{sec:euclid} 

The science goal of this citizen science experiment (available at \url{https://www.zooniverse.org/projects/hughdickinson/euclid-challenge-the-machines}) was to see whether human volunteer classifiers can find strong gravitational lens events better than machine learning. The Euclid consortium had recently ran a Strong Gravitational Lensing Challenge (with challenge data deposited in the public domain), in which teams contributed machine learning algorithms to find lensing events in simulated imaging data from single-band Euclid and the multi-band ground-based KiDS (Kilo Degree Survey) project. One valiant expert volunteer visually classified the entire $\sim100\,000$ image set. One of the great surprises of this challenge was that the valiant expert did not win the challenge \cite{Metcalf+19}. Several machine learning algorithms were more effective than the human expert in finding lenses. But this is a feature recognition problem in which a non-expert can be very well trained, so the question is whether the artificial intelligence has exceeded human abilities or whether some humans can still beat the machine. Since we were more likely to engage volunteers with colourful images, we selected the simulated KiDS data for crowdsourced analysis, rather than that of Euclid. 

Our tactic for educational resources associated with the mass participation experiments was to embed the material directly into the citizen science workflow. This means that the participant volunteers can be introduced gently to the science context as their involvement in the project extends.

We selected this challenge data for several reasons:
\begin{itemize}
\item It was to be the first of several similar  challenges, so new versions of the crowdsourcing experiment could be run at later dates.
\item There is a precedent for successful crowdsourcing in strong gravitational lensing, in the Space Warps experiment.
\item Similar morphological searches for strong gravitational lenses can also be made in other ASTERICS and ASTERICS-related facilities, such as the SKA, the Large Synoptic Survey Telescope, the Hubble Space Telescope, and others.
\end{itemize}
Besides the particular science objectives of the experiment, we were also aiming to find experiments that are in some way aligned with the primary science goals of the facility, and which are likely to have some longer-term traction or application to other facilities. In the case of Euclid, the enormous hundred-fold increase in known strong gravitational lensed systems that Euclid will bring will provide independent statistical constraints on dark energy parameters.

The drafting of the experiment itself and its associated educational resources was done very quickly and efficiently on the Zooniverse's Panoptes platform. We helped in the creation of the new mobile app functionality for Zooniverse, using a ``Tinder''-like swiping left and right to indicate a binary classification. This citizen science experiment was one of the first to pioneer this mobile citizen science interface.

However, that is not to say that all aspects of this project were trivial to create. The classification data set was examined carefully by ourselves and by our collaborators in the Zooniverse Space Warps project. We discovered that while most of the simulated data sets were representative of the known gravitational lens systems discovered in the COSMOS survey by the Hubble Space Telescope, there was also a significant subset of outlier systems that appeared to be unphysical. Typically, these were systems in which there was an unusually faint foreground lensing galaxy with the background galaxy warped into a wide Einstein ring. Since the ring radius is a reliable proxy for the total mass of the system, this would mean that the simulated galaxies either have an unphysical mass-to-light ratio or have been placed at unphysically high redshifts (i.e. greater distances).
We therefore took care to pre-filter the data set to be classified, in order to better reflect the observed parameter distributions in the known Hubble Space Telescope gravitational lenses. This process took several months. It should be added that none of these anomalies that we discovered had been uncovered by any of the AI algorithms or human inspections of the data set prior to our work. Further detail of the results can be found in \cite{DaviesPhD}.

\subsection{Galaxy Zoo: Clump Scout}\label{sec:galaxyzooclumpscout}
One of the main goals for modern observational cosmology is to discover and understand how galaxies and their constituent substructures have assembled and evolved throughout cosmic history. The diverse observed morphologies of individual galaxies are not only indicative of their current composition, but also encode a detailed record of their assembly histories, their past and ongoing star formation, and their interaction with local environments.  
Galaxies grow by forming stars. Today, the Hubble Space Telescope can detect distinct star-forming structures inside the galaxies that populated the Universe when it was less than a quarter of its current age. These early galaxies look very different to their modern-day counterparts. Their disks are thick, turbulent and violent environments, where hundreds of new stars are born every year. Many also exhibit giant regions of enhanced star formation that appear as bright clumps in telescope images. In contrast, today's star forming galaxies are typically much more placid. Their disks are thin and well-ordered and clumpy star formation is much less common.

These profound differences raise obvious questions. Which physical mechanisms drove the observed evolution in star formation activity? Why are giant star forming clumps so much more common in the early Universe?  

To understand why clumpy galaxies became so rare, we need to find and investigate as many examples as possible. One potential approach involves training modern deep learning algorithms that use deep learning to identify galaxies with clumps. However, appropriately labelled training data for clump detection is scarce and laborious to generate. Moreover, automatic algorithms struggle to operate effectively if their limited training datasets underrepresent the diversity of the data being analysed. In contrast, human beings working in collaboration can extrapolate successfully from a handful of examples.

To benefit from this impressive human capability, we used the Zooniverse platform to develop a new citizen science project called Galaxy Zoo: Clump Scout. The project invites the general public to examine images of galaxies obtained by the Sloan Digital Sky Survey (SDSS) and annotate all the clumps they can see. By participating in Galaxy Zoo: Clump Scout, volunteer clicks will identify the locations of clumps within thousands of galaxies in the nearby Universe. The project uses a novel Bayesian aggregation algorithm that dynamically derives a consensus for the clump locations based on the annotations provided by multiple volunteers for the same image. The algorithm also estimates the reliability of the dynamic consensus, which helps to ensure completeness while avoiding spurious clump detections. Galaxy Zoo: Clump Scout represents one of the first large-scale studies of clumps in local galaxies.  

In the future, new space telescopes like Euclid will image more than a billion galaxies. Using citizen science to manually check so many galaxies for clumps would take many years, even for the most dedicated Clump Scout volunteers. The speed of computer algorithms will be required to process such large volumes of data and so we have trained Deep Learning models for clump detection and localisation based on the Faster RCNN architecture that exhibit very good performance on a wide range of galaxy images. However, there will always be galaxy images that confuse the computer algorithms and we will need the help of the Clump Scout volunteers to step in when deep learning fails. Even more importantly, human beings inspecting images are much better at spotting any unusual or unexpected phenomena that single-minded algorithms would just ignore. Indeed, the history of citizen science is full of examples when keen-eyed volunteers make serendipitous discoveries. Projects such as Clump Scout will help to maintain this tradition in the future.

The Galaxy Zoo: Clump Scout project has now concluded after collecting 1,738,822 from 13,762 volunteer citizen scientists who collectively annotated the locations of visible giant star-forming clumps in 85,286 low-redshift galaxies. To date, this project has generated two refereed journal papers. Dickinson et al. \cite{Dickinson2022} 
presented a novel algorithm for aggregation of two-dimensional citizen science data, deployed on the clump annotations in this project. Adams et al. \cite{Adams2022}
presented the first results of the observed clump fraction in galaxies from this project, revealing a sharp decline in the clumpy fraction from redshift 0.5 to the present day. The aggregated annotations have since been used as truth sets for training machine learning algorithms, which will be the subject of future papers. 

\subsection{SuperWASP: Black Hole Hunters}\label{sec:superwaspblackholehunters}
SuperWASP: Black Hole Hunters (SWBHH, \url{https://www.zooniverse.org/projects/hughdickinson/superwasp-black-hole-hunters}) was a new citizen science project that has volunteers search for an extremely rare event, that of a hidden black hole orbiting a normal star in a binary system. 

To fully understand the lifecycle of stars and how they contribute to the properties of a galaxy, astronomers not only look at regular stars, but also the remnants they become at the end of their lives: black holes and neutron stars. These compact objects form in the violent supernova explosions of massive dying stars. However, most stars have a companion (a binary system), so provided this explosion does not tear apart the system, then the result is the companion star and the compact object orbiting each other. A couple of hundred compact objects have so far been discovered because their orbit has decayed inwards to the point that they can accrete matter from the companion star. However, simulations predict there should be hundreds of thousands of star-compact object binary systems, so another method is needed to identify these elusive objects.

One promising method is that of gravitational microlensing. When a foreground object such as a black hole or neutron star passes in front of a distant background star, the foreground object can act as a lens for the background star’s light, causing the light to bend around it. Provided high enough resolution imaging, we observe this as the background star being distorted into a ring of light (called an Einstein ring). Unfortunately, for small objects like black holes and neutron stars, the angular resolution limits of telescopes prevent us from seeing these distortions directly. However, it is possible to infer these distortions by observing a temporary increase in brightness as the object passes between us and the background star. Such a microlensing event can even occur for the aforementioned star-compact object binary systems, with the compact object passing between us and its companion star. This produces a characteristic periodic increase in brightness that we can use to detect the compact object, and even learn about its properties, but to do this requires us to view the system almost perfectly edge-on.

These ‘self-lensing’ signals are expected to be extremely rare - so far, nobody has seen conclusive evidence of a self-lensing signal from a black hole. Even though simulations predict numerous hidden black holes in our galaxy, only a handful will be orbiting in a plane that means they pass in front of their companion star when viewed from the Earth. To find such rare and infrequent events, we need to use survey data that covers large regions of the sky and spans long periods of time.

SuperWASP was an experiment to monitor the sky in a search for exoplanets. SuperWASP’s exploration of the time domain parameter space in astronomy is a direct precursor to the Legacy Survey of Space and Time on the Vera Rubin Observatory, one of the forthcoming ESFRIs supported by ESCAPE. The now-decommissioned SuperWASP telescopes are also an excellent instrument to use for self-lensing searches. During its operational lifetime SuperWASP accumulated light curve data for millions of stars. It observed a very large fraction of the sky over a period of approximately 8 years with as little as 40 minutes between repeated observations. The primary shortcoming of SuperWASP data is that they are very noisy, which complicates automatic detection of subtle self-lensing signals and necessitates manual inspection of lightcurves.

To enable manual inspection of so many lightcurves, we designed and deployed SWBHH as a citizen science project on the Zooniverse web platform. SWBHH engaged 5\,673 volunteers who have collectively completed a spectacular 2.1 million inspections of all its 208\,700 lightcurves. The project launch was timed to coincide with the launch of a prime-time BBC science series “Universe”, which was presented by the popular UK-based science communicator, Professor Brian Cox. The SWBHH project was promoted to viewers of “Universe” using an accompanying poster designed by members of the ESCAPE collaboration. The SWBHH project was also promoted by Serjeant and Dickinson as part of an online panel discussion about black holes during British Science Week 2022 (\url{https://www.superwasp.org/britishscienceweek/}). During the event alone, the audience completed over 7\,000 classifications and the total for the day was over 14\,000 classifications. 

\subsection{Galaxy Zoo: Cosmic Dawn}\label{sec:galaxyzoocosmicdawn}
Galaxy Zoo: Cosmic Dawn was a new citizen science project that formed the latest iteration of the longest running project on the Zooniverse, Galaxy Zoo, which aims to classify images of galaxies based on their visual morphologies. 

A core aim of extragalactic astronomy is to study how galaxies form and evolve over cosmic time, including understanding the physical mechanisms that govern their structures and produce the wide range of galaxies we observe. Galaxies have a variety of shapes, from ball-like ellipticals to those with grand spiral arms, and different colours that indicate the presence and composition of dust and stars of different ages. These properties pertain to that galaxy’s history, including through its rate of star formation, the activity of its central black hole, and the merging of it with other galaxies over cosmological time. Galaxies also typically live within extensive dark matter haloes that can, in rare circumstances, act as a (gravitational) lens to distort and magnify the light of distant background galaxies that would otherwise be hidden from us. Additionally, low surface brightness (LSB) galaxies contain a higher ratio of dark matter to baryonic matter, making them useful probes for studying the impact of dark matter, but are inherently faint and therefore difficult to detect. Hence, the appearance of a galaxy is the result of a combination of properties, and to study the impacts of these properties requires large samples of classified galaxies. Modern surveys are capable of collecting such large samples for fainter objects using deeper imaging, moving towards higher redshifts in order to examine galaxies from further back in time and their evolution from then to now. Any of those identified as a lens would additionally allow for the study of even more distant galaxies from the earliest epochs of galaxy formation.

The Cosmic Dawn survey aims to understand how galaxies, black holes and dark matter haloes co-evolve from the epoch of reionization (when the Universe was around 500 million years old) to the present. As a 50 square degree multi-wavelength survey of the Euclid deep fields, it covers some of the darkest areas of the sky that have been selected for study by multiple international observatories, partly in preparation for the ultra-deep photometry and spectroscopy to be produced by the upcoming Euclid mission. These deep fields allow for the study of large numbers of galaxies going back to when the first of them formed within the first billion years after the Big Bang.

One of the programs forming Cosmic Dawn is the Hawaii Two-0 (H20) survey of the Euclid deep calibration fields, which involves deep multi-band imaging of the 10 square degree Euclid Deep Field North (EDF-N) using the Hyper Suprime-Cam (HSC) on the 8.2m Subaru Telescope on the summit of Mauna Kea in Hawaii. The ultra-deep HSC imaging contains around a million galaxies per square degree down to their magnitude limit, with data processing still ongoing. Through combining this with Spitzer Space Telescope infrared imaging and Keck DEIMOS (DEep Imaging Multi-Object Spectrograph) spectroscopy, H20 will advance the study of galaxy evolution and co-evolution with dark matter haloes out to a high redshift of z = 7 ($<$800 million years since the Big Bang), pushing the boundaries of extragalactic astronomy.

Citizen science helps play a crucial role in the examination of such large data sets, with volunteers able to classify huge amounts of data much faster than a small team of researchers. While machine learning methods has developed rapidly over the past few years, volunteers still outmatch them when it comes to classifying complex features or the serendipitous discovery of rare objects. As such, the Galaxy Zoo: Cosmic Dawn project has volunteers classify the hundreds of thousands of galaxies detected in H20’s HSC imaging of the EDF-N, exploring a region of intense study with deep multiband imaging. These classified images extend to higher redshift sources and provide a means of statistically studying objects such as galaxies with LSB features while expanding the list of interesting, rarer objects found through serendipitous discovery. In addition, mapping the EDF-N is important for future surveys such as the Euclid mission, with classifications providing a basis for rapid follow-up of the most interesting objects. HSC is also a precursor for the Legacy Survey of Space and Time (LSST) Vera C. Rubin Observatory, with this classification work providing multiband ground truth sets for use in training deep learning models, such as for strong gravitational lens finding.

The project workflow involved presenting colour postage stamp cutout images of each galaxy to volunteers to classify following the set of questions developed by Galaxy Zoo, modified to suit these HSC images. Initially focusing on 2 square degrees in the EDF-N, it can be extended to a wider area as the processed H20 data becomes available. Due to the images featuring sources at higher redshifts than is typical for Galaxy Zoo, a new colour scaling had been implemented. Additionally, the added confusion noise from the increased number of background galaxies in these deeper images proved challenging for the H20 team’s source detection and modelling, hence requiring re-sizing of their cutouts to correctly display sources. The Zooniverse tutorial and help text for the project were updated to reflect this and explain the project’s new data, and the “star/artefact” option normally presented in the Galaxy Zoo workflow was also  expanded in order to aid the H20 team with improving their source detection and modelling methods going forward.

When the project had its public launch, a blog post was announced alongside it informing volunteers of the launch and to keep an eye out for and tag any extremely red sources or gravitational lenses due to their rarity. This also featured in the tutorial and help sections of the project, and, as volunteers can select lensing features as part of the workflow, this additionally allowed the Galaxy Zoo team to investigate how tagging rare sources compares to asking via the workflow questions. The Galaxy Zoo: Cosmic Dawn project was completed in June 2023, and a general data release paper is currently in production. The volunteers’ classifications will also serve to generate multiple publications, including the detection of strong gravitational lenses, statistics of clumpy galaxies, and examinations of galaxies with low surface brightness features.

\subsection{Radio Galaxy Zoo: LOFAR}\label{sec:radiogalaxyzoolofar} 
The Low Frequency Array (LOFAR) is a large interferometric array of radio telescopes located primarily in the Netherlands, but with outlying antennae dispersed across Europe. LOFAR is also a recognised science and technology pathfinder facility for the next-generation radio telescope, the Square Kilometre Array (SKA).  

Radio Galaxy Zoo: LOFAR (\url{https://www.zooniverse.org/projects/chrismrp/radio-galaxy-zoo-lofar}) was a new citizen science project led by ASTRON in the Netherlands with substantial ESCAPE-funded support provided by the Zooniverse platform and the Open University (OU). The project invites volunteers to classify radio images extracted from the first data release of the LOFAR Two-metre Sky Survey (LoTSS) which covers 424 square degrees in the region of the HETDEX Spring Field. In this release, 325,694 individual radio sources were detected with a signal five times greater than a typical background noise fluctuation.

Classification entails attribution of distinct regions of radio emission to a single origin and (where possible) identifying an optical counterpart for the radio emission’s source. By Zooniverse standards this is a very complicated analysis task, which requires consideration of multiple images, representing radio and optical data. Moreover, the degree of scientific comprehension that volunteers require to successfully provide the required classifications is more than typical Zooniverse projects, which often rely on somewhat mechanical “microtasks” that can be performed without complete understanding.

To render such complex classifications tractable for citizen scientists, the OU and Zooniverse teams have developed an advanced volunteer training and feedback system. The project uses a tutorial video paired with a dedicated training workflow that allows volunteers to mimic the classification process as demonstrated by one of the LOFAR project scientists. The training workflow presents subjects in the same order as they appear in the video (unlike the normal random ordering employed by the Zooniverse platform) and volunteers receive real-time feedback in response to the annotations they provide. This is the most advanced training infrastructure that has been deployed using the Zooniverse project builder platform and the upgrades that have been developed by the OU and Zooniverse with ESCAPE support will be available for future CS projects to leverage. It has been shown that volunteers’ confidence is a critical factor in citizen science projects, which improves classification accuracy and volunteer retention. 

The project is now complete, with 11,583 volunteers having contributed 0.95 million classifications on a total of 189,375 subjects. A refereed journal paper is now in preparation on the catalogue of optical identifications generated in this project. 

\subsection{Knitting Patterns}\label{sec:knittingpatterns}
Our ambition has always been to extend the EOSC citizen science demonstrators to beyond the subject specialist areas of the ESCAPE EOSC cell. Therefore, with the support of the EOSC-Future project, we have assisted in the development of the Knitting Leaflets citizen science project (\url{https://www.zooniverse.org/projects/elliereed185/knitting-leaflet-project}). This project aims to discover more about how knitting and knitwear developed in Britain during the twentieth century, by carefully characterising information on the front covers of knitting pattern leaflets and magazines, to yield knowledge of how versions of femininity connected to fashion and consumption, and how they changed over time. Three workflows were developed requiring increasing levels of knowledge in this subject area. The project had its public launch on 27th September 2023, and proved exceedingly popular as it was completed within a week, with 1\,327 volunteers making a total of 52\,928 classifications of 9\,624 subjects.

The Knitting Leaflets project was promoted to the Zooniverse community of over 2.6 million registered users through the Zooniverse front page as well as through the Zooniverse email newsletter (announcements@lists.zooniverse.org). It was also promoted in an article on the Knitting Industry website (\url{https://www.knittingindustry.com/creative/the-knitting-leaflet-project/}; 400+ listed views) as well as the Knitting History Forum Facebook page (\url{https://www.facebook.com/KnittingHistoryForum/?locale=en_GB}; 1\,600 followers), and the project also featured in the forums of the Gardeners’ World website (\url{https://forum.gardenersworld.com/discussion/1077606/anybody-interested-in-the-history-of-knitting-patterns}).

\subsection{African Indigenous Knowledge}\label{sec:aik}
African Indigenous Knowledge (AIK; \url{https://www.zooniverse.org/projects/xsr/african-indigenous-knowledge-aik-m}) is the second citizen science project we have helped to develop to extend the EOSC citizen science demonstrators beyond the ESCAPE remit, with the project also using the Zooniverse platform. This project is still under development, but aims to study 
traditional African indigenous food system knowledge for 
better
food production, processing and consumption in Africa. The project asks volunteers to classify the types of food and tools present in photographs from traditional farmers in Sierra Leone, and is currently paused while data collection is underway.

\section{Discussion}
\subsection{Tools and Data}\label{sec:tools}
Subject-specialist researchers must often go to substantial lengths to convert experimental data into high-level “subjects” (images, videos etc.) that can be successfully interpreted and analysed by non-expert volunteers. Therefore, in order to streamline this process and reduce the workload of running a successful citizen science project, we have created tools and template workflows that generate attractive subject data, upload them to the Zooniverse servers and manage them effectively as the volunteer classifications accumulate.

Firstly, we created a Jupyter notebook and documentary materials demonstrating web-interface based and programmatic (scriptable) Zooniverse project management including project and workflow creation, subject creation and upload, improving the volunteer experience through extra metadata, retrieving a list of subject sets, downloading classification and subject data, setting up volunteer feedback and training. This tutorial makes use of example material (subjects, metadata, classifications) from the SuperWASP Variable Stars Zooniverse project, which in turn was created as an EOSC demonstrator, and which is also discussed in section \ref{sec:superwaspvariablestars} above. The notebook is available at \url{https://git.astron.nl/astron-sdc/escape-wp5/workflows/zooniverse-advanced-project-building}, and a recorded walkthrough of this advanced tutorial is available at \url{https://youtu.be/o9SzgsZvOCg}~. 

The ESAP Archives (accessible via the ESAP GUI) include data retrieval from the Zooniverse Classification Database using the ESAP Shopping Basket. We therefore also created a tutorial on loading Zooniverse data from a saved shopping basket into a notebook and performing simple aggregation of the classification results (also available as an Interactive Analysis workflow). The tutorial notebook includes importing the Panoptes Python Client and ESAP client Zooniverse connector, retrieving items added to an ESAP Shopping Basket, and an example of retrieval and analysis of data from the Muon Hunter Zooniverse project, which in turn in turn was created as a EOSC demonstrator, and which is also discussed in section \ref{sec:muonhunter} above. The notebook covers the downloading and processing of classification data, as well as plotting the data and performing simple aggregation of the results. The tutorial is available at \url{https://git.astron.nl/astron-sdc/escape-wp5/workflows/muon-hunters-example/}~.  

We also created a Jupyter notebook tutorial and documentary materials demonstrating how to set up an active learning framework to continuously train machine learning models using volunteer classifications of optimally selected subjects. Zooniverse's Caesar advanced retirement and aggregation engine allows for the setup of more advanced rules for retiring subjects (as opposed to the default method of retiring after that subject has been classified a certain number of times). Caesar also provides a powerful way of collecting and analysing volunteer classifications (aggregation). This tutorial makes use of example material (subjects, metadata, classifications) from the Penguin Watch Zooniverse project, which involves counting the numbers of penguin adults, chicks and eggs in images to help understand their lives and environment. A recorded walkthrough of this advanced tutorial is available at \url{https://youtu.be/o9SzgsZvOCg?t=3840} while the notebook itself is at \url{https://git.astron.nl/astron-sdc/escape-wp5/workflows/zooniverse-advanced-aggregation-with-caesar}~. 

Finally, we created a notebook demonstrating how to integrate Zooniverse projects with existing machine learning frameworks and combine volunteer classifications with machine learning predictions. The tutorial introduces advanced retirement rules using machine learning, with choices of either pre-classifying with machine learning in preparation for volunteer classifications, or performing ``on the fly'' retirement decisions made after both machine learning and volunteer classifications. The tutorial shows how to use machine learning to filter out uninteresting subjects prior to volunteer classification, and demonstrates setting up active learning in which volunteer classifications train the machine learning model, which in turn handles the ``boring'' subjects and leaves the more challenging/interesting subjects for volunteers. The notebook is available at \url{https://git.astron.nl/astron-sdc/escape-wp5/workflows/zooniverse-integrating-machine-learning} and a recorded walkthrough is at \url{https://youtu.be/o9SzgsZvOCg?t=8218}\,, conducted as part of our first ESCAPE Citizen Science Workshop.

\subsection{Ambition for the future}\label{sec:vision}
\begin{figure*}
\includegraphics[width=\textwidth]{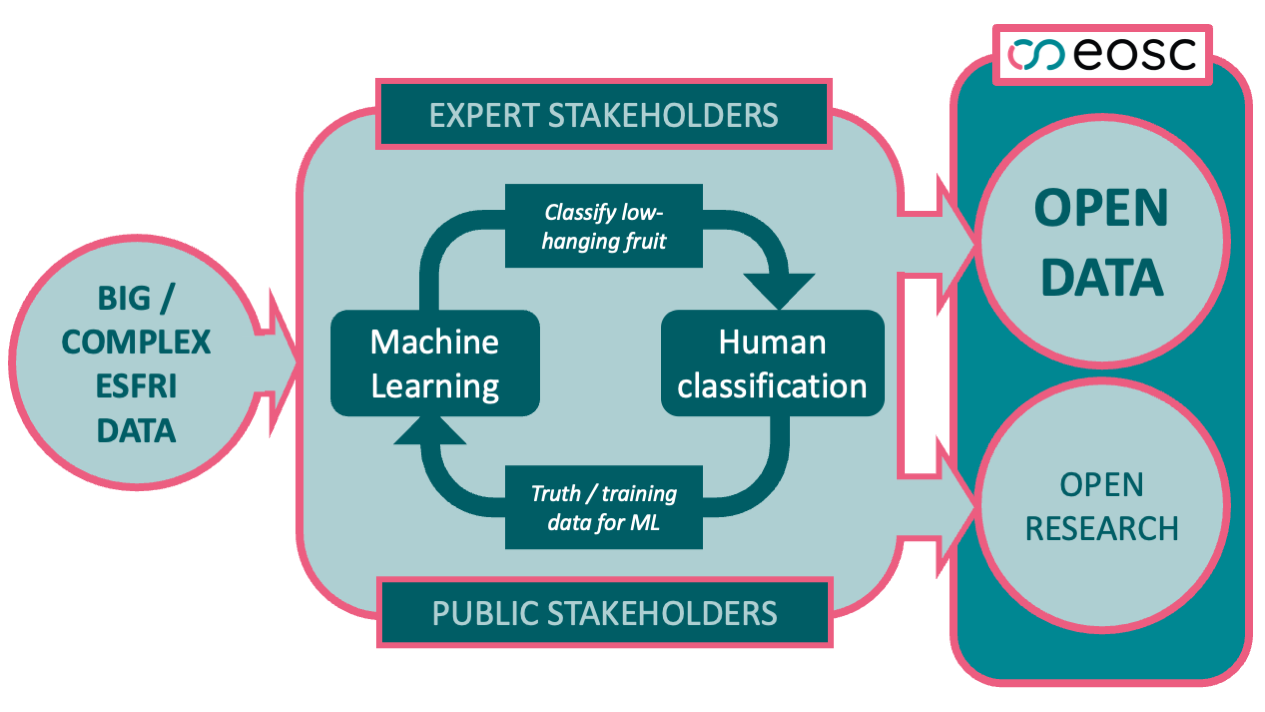}
\caption{\label{fig:cs_schematic} Schematic PERT diagram showing selected workflows of 
big/complex data from the European Strategic Forum for Research Infrastructures (ESFRIs),  
and the external context of the European Open Science Cloud (EOSC). This illustrative workflow shows the virtuous circle between human and machine learning.} 
\end{figure*}

Our vision is for the community to have access to clear exemplars of planning, creating and managing crowdsourced data mining on EOSC, implementing machine learning in real time, across a wide range of scientific domains, which they can use as templates to deploy with ease. Through these activities, our vision is to increase the size of the community making real scientific engagement with EOSC by orders of magnitude, solving the difficult problem of usefulness of FAIR data by giving non-specialists a carefully curated and educationally supportive experience of EOSC. The workflow is illustrated schematically in Fig.\,\ref{fig:cs_schematic}. 

In realising this EOSC citizen science vision, there are some important lessons from ESCAPE and other projects. One is that interoperability between citizen science services is not automatic; rather, it requires (funded) developer resource to create the interoperability, as was the case in the Zooniverse/CitSci.org cross-integration. Secondly, tutorials, notebook resources and other training materials also need funding for their creation. Thirdly, even with abundant online training materials, the most effective intervention is to have a “support scientist” on hand to assist in the creation, operation and ultimately the science exploitation of citizen science projects. Finally, all the citizen science demonstrators created in ESCAPE were driven by fundamental science questions, and had active commitments from the subject-specialist science teams who were motivated to find answers to these fundamental science questions. Science team engagement with volunteers is one of the key factors influencing success in citizen science projects, and all the ESCAPE citizen science demonstrators have been objectively extremely successful, but the lesson in EOSC is deeper: technical developments are, and must be, precipitated bottom-up by the science use cases. Features that are “nice to have” are not built unless they also have science drivers. This is consistently true in citizen science contexts such as the Zooniverse, and arguably also should apply more widely to EOSC development in general. 

With those lessons in mind, we aspire to closer cooperation of citizen science platforms with EOSC. Our vision is to increase citizen science integration with EOSC services, expanding citizen science domains through EOSC. Partly, this will require advocates skilled in using citizen science platforms to engage with wider subject-specialist communities, possibly (though not necessarily) through the creation of multi-disciplinary demonstrator projects driven by specific science needs from those communities. It may also help to train, or to simply provide as a service, the data visualisation tools that are widely used in the ESCAPE science domain areas but less common elsewhere. We would also support the creation of a Citizen Science Task Force within EOSC communities such as the EOSC Association. 

We also aspire to deeper technology integration. Citizen science volunteers in the astrophysics domain already routinely ask or expect the functionality to navigate from an astronomical image subject in Zooniverse, out to a virtual observatory interface to interrogate other multi-wavelength data or to explore the data to be classified with a greater flexibility that is available on the Zooniverse platform alone, all in aid of open-ended and curiosity-driven enquiry by volunteers. For technical reasons relating to the underlying technologies, it is not simple to integrate virtual observatory tools directly into the Zooniverse platform itself, but the links out and in again can be made as frictionless as possible for particular science cases. Again, developments must be driven by community need, whether those communities are subject specialists or the science-inclined public. 

One important caveat to this deeper technology integration is that astronomy, astroparticle physics and particle physics are relatively benign subject areas for non-specialists to explore data, in that there are not many serious consequences for the misunderstanding and misuse of data. This is not always the case in other domains, such as healthcare or climate science, where the policy stakes are much higher and there are even efforts by malicious actors to disseminate disinformation. The wider integration of EOSC technologies with citizen science must therefore account for the wider policy and ethics landscapes of each science domain\footnote{Further discussion of the policy environment for citizen science can be found in Angelidakis et al., this volume.}.
During the course of the ESCAPE project there has been a growing interest in science analysis platforms as a whole, not just in ESCAPE’s own ESFRIs Science Analysis Platform (ESAP), but also in the Rubin observatory science analysis platform, and in the European Space Agency’s new Datalabs platform. As part of the efforts of these platforms towards interoperability within the EOSC ecosystem, the citizen science use cases need to be included explicitly. With a suitable funding stream for the server and for compute, it may also be beneficial to run a dedicated ESAP instance for citizen science, if the user communities do not have routine access to these other science analysis platforms. 

We would also like to see greater development of open data standards in citizen science, including creating a registry of terminology, models/formats, defining reporting guidelines, describing data policies, and creating models for identifier schemas related to citizen science. 

Finally, as machine learning technologies improve, the synergies and opportunities with crowdsourced data mining will evolve. This may require more science-driven multi-disciplinary worked examples of plug-and-play citizen science notebooks with embedded machine learning integration, to facilitate citizen science uptake throughout the EOSC science domains. 

\subsection{Could you use citizen science for your physical science research?}\label{sec:motivate}
We conclude this review with some brief reflections on the types of physical systems and problems that are amenable to citizen science approaches. The most important lesson from the examples above is that every case has been driven by the fundamental science question. In no cases has the activity been driven primarily by a desire for public engagement {\referee or outreach}. In fact, it is logically impossible to optimise against more than one parameter. For example, one can seek to climb the highest mountain on Earth (Mt. Everest), or the hardest to climb (arguably K2), but it is not possible to simultaneously choose the highest {\it and} hardest. One can optimise subject to a constraint, such as maximising the science return given the use of crowdsourcing from volunteers or getting to the highest point in a country subject to the constraint that it is accessible by road, but there must always be only one parameter to optimise. Therefore we do not make recommendations for public engagement {\referee or outreach} in this review, even though there is abundant evidence that citizen science can demonstrably benefit public engagement, {\referee outreach} and societal impact {\referee (see the outreach and societal impact studies in} e.g. \cite{Prather2013}, \cite{LuczakRoesch2014}, \cite{Masters+16}, \cite{Trouille+19}). If public engagement is the primary goal, then rather than using techniques optimised for something else, in many cases it may be more effective to make more targeted interventions to support that primary goal; {\referee in other words, if the most important factor to you is that you want to do outreach, then you may well be better off optimising your efforts to that specific goal, rather than deploying citizen science activities optimised for something else}. We also do not address techniques of co-creation and the use of volunteers throughout the research cycle (e.g. \cite{SenabreHidalgo2021}) since, with very few exceptions, this methodology is unsuitable to the physical sciences because of the depth and breadth of background information necessary. Even in the exceptions that exist, the volunteer involvement has to be carefully managed (e.g. \cite{Keel+22}).

As a useful heuristic it is worth considering the motivations for Zooniverse volunteers\footnote{These five insights are due originally to Becky Smethurst.}:
\begin{itemize}
\item {\bf Is it beautiful?} In other words, is the project visually engaging? For example, part of the success of the Muon Hunter project (section \ref{sec:muonhunter}) can be attributed to the strikingly pretty colourful arcs to be identified. 
\item {\bf How easy is it?} Simpler workflows will engage larger cohorts of volunteers. Conversely, complex analysis will draw smaller but more dedicated volunteer teams. At the simplest end, binary classifications can be deployed on the Zooniverse mobile platform, as in section \ref{sec:euclid}. 
\item {\bf How important is it?} The volunteers may be spending a large fraction of their free time on the project, so the onus is on the science team to convince the volunteers that the task is worthwhile and important. 
\item {\bf How much am I learning?} The in-built tutorial material supports volunteer learning, and although the learning journey is not the primary objective of the Zooniverse research projects, the projects that provide a sense of personal enrichments are more likely to succeed in practice. Moreover, direct science team engagement on online talk forums is essential not just for fostering volunteer engagement, but also for assessing the progress of the project and discussing volunteer discoveries. Exactly this science team engagement with volunteers led to the successful but unexpected discoveries of the new classes of Voorwerpjes \cite{Lintott+09} and green pea galaxies \cite{Cardamone+09}, identified first by volunteers in Galaxy Zoo.
\item {\bf How famous could I get?} Some citizen science projects have a tantalising possibility of a spectacular breakthrough, and this on its own can sometimes be enough to motivate volunteers. An analogy might be made with a lottery: time or money is spent on entertainment, with the tantalising possibility of a large return. In the context of citizen science volunteers it is also worth considering (or just asking) what the volunteers would value as a recognition, which may vary from project to project. For example, while the science team might place the highest value on a co-authorship of a high impact paper, a volunteer might place a higher value on a name appearing on a NASA website. 
\end{itemize}
{\referee This is not an exhaustive list of volunteer motivations. Indeed for some citizen science projects, competition may be a factor. We will discuss gamification use cases below, which we use to refer to short-term competition for a particular task or for more long-term achievements on e.g. ranking on a leaderboard.}

This review has focused mainly on imaging use cases in which subjective human assessments are used, whether for image annotation or image classification, which in turn can be used as training sets for machine learning technologies. These machines can then be used to ``mop up'' the easiest classification subjects, allowing human effort to be refocused on more difficult cases, ultimately creating a virtuous circle between human and machine learning (Fig.\,\ref{fig:cs_schematic}). This on its own captures a very wide variety of research use cases in the physical sciences, but there are many other ways that human effort can be deployed in the physical sciences. 

One immediate generalisation is that human volunteers can parse more than visual data. Several efforts have already been made at sonifying astronomical data (e.g. \cite{SonoUno}, \cite{Zanella+22}) and there have been several successful Zooniverse project that require volunteers to parse audio, including assessing the maturity of language development in babies (e.g. \cite{babysounds1}, \cite{babysounds2}) or identifying urban noise in New York with the help of a spectrogram linked to audio (e.g. \cite{SoundsOfNewYork}). In the latter, the spectrograms presented to volunteers are qualitatively similar to those in gravitational wave Zooniverse projects. Perhaps because the professional community is not yet accustomed to parsing their own data through audio, this currently remains an under-explored use case for crowdsourced data mining in the physical sciences. 

Going beyond the use of volunteers as distributed, massively parallel biological computers for data mining, there are many other ways in which non-specialists can contribute to research in the physical sciences. Astronomy in particular has a distinguished history of pioneering crowdsourced data collection (e.g. \cite{AllanChapman}), from the discovery of supernovae and comets to the identification of high energy cosmic rays (section \ref{sec:credo}) to the discovery of meteors at radio and optical wavelengths (e.g. \cite{Rendtel17}) and of meteorites (e.g. \cite{King+22}, \cite{OBrien+22}). The Search for Extraterrestrial Intelligence (SETI) pioneered the deployment of otherwise idle CPU on volunteer computers for data analysis, in the SETI@Home screensaver. More opportunities in the physical sciences may arise as the balance evolves between the availability of high performance computing for research, and the potential crowdsourced capabilities of volunteer home computing.

``Amateur'' astronomy includes volunteers who make very significant personal financial investments in hardware for observational astronomy, to the extent that there is an overlap between the capabilities of the smallest professional facilities and the largest amateur ones, at around $\sim0.5$\,m diameter primary mirrors for optical telescopes. This makes it possible for the non-professional community to contribute directly to monitoring campaigns of bright targets, such as measuring transiting exoplanets or monitoring nearby supernova light curves. Moreover, the extremely distributed nature of the data collection can be a decisive advantage for some use cases, such as in multi-messenger astronomy, where the detection of an optical transient may be prevented at one location by local weather conditions but possible at another. Even with the advent of large sky monitoring projects such as the Legacy Survey of Space and Time at the Vera Rubin Observatory, there will always be multi-messenger events inaccessible to any particular facility and so some potential for non-professional contributions will continue. The International Astronomical Union has a working group dedicated to promoting and facilitating research between professional and non-professional communities, at \url{https://www.iau.org/science/scientific_bodies/working_groups/professional-amateur/}~.

Some citizen science projects have explored other approaches to motivating volunteer effort, such as gamification and contests. The FoldIt project seeks volunteer effort to compete to find novel protein folding configurations \cite{Koepnick2019}, a problem well-known to be NP-hard \cite{Fraenkel1993}. 
Another computational problem that is NP-hard is to find the longest common subsequence among sequences; this is relevant to the analysis of DNA, so there are other gamified citizen science applications in bioinformatics such as the Phylo project (\url{https://phylo.cs.mcgill.ca/}) and Borderlands (\url{https://borderlands.2k.com/news/borderlands-science/}). There are arguably comparably difficult NP-hard problems in the physical sciences, though so far gamification has been attempted for computationally simpler problems, such as exoplanet discovery in Project Discover, with volunteers competing for rankings in accuracy. There may be underexplored opportunities in the physical sciences for volunteer competitions in harder problems.

\bmhead{Acknowledgments}
\hspace*{0cm}
{\referee We thank the anonymous referee for helpful suggestions.}
The authors were part-funded by the ASTERICS, ESCAPE and EOSC-Future projects, through a variety of European Commission Horizon calls. 
ASTERICS is a project supported by the European Commission Framework Programme Horizon 2020 Research and Innovation action under grant agreement no. 653477.
ESCAPE - The European Science Cluster of Astronomy and Particle Physics ESFRI Research Infrastructures has received funding from the European Union’s Horizon 2020 research and innovation programme under Grant Agreement no. 824064. 
The EOSC Future project is co-funded by the European Union Horizon Programme call INFRAEOSC-03-2020 - Grant Agreement Number 101017536.



\section*{Declarations}
The authors declare no competing interests. Funding sources are declared above. All of the software notebooks descrbed in this paper are part of the ESCAPE Open Source Software Repository (OSSR). All of the ESCAPE OSSR content is hosted in the Zenodo ESCAPE2020 community, at \hyperlink{https://zenodo.org/communities/escape2020}{https://zenodo.org/communities/escape2020}~~.

\section*{Data availability statement}
No Data associated in the manuscript.

\bibliography{sn-bibliography.bib}


\begin{thebibliography}{37}
\ifx \bisbn   \undefined \def \bisbn  #1{ISBN #1}\fi
\ifx \binits  \undefined \def \binits#1{#1}\fi
\ifx \bauthor  \undefined \def \bauthor#1{#1}\fi
\ifx \batitle  \undefined \def \batitle#1{#1}\fi
\ifx \bjtitle  \undefined \def \bjtitle#1{#1}\fi
\ifx \bvolume  \undefined \def \bvolume#1{\textbf{#1}}\fi
\ifx \byear  \undefined \def \byear#1{#1}\fi
\ifx \bissue  \undefined \def \bissue#1{#1}\fi
\ifx \bfpage  \undefined \def \bfpage#1{#1}\fi
\ifx \blpage  \undefined \def \blpage #1{#1}\fi
\ifx \burl  \undefined \def \burl#1{\textsf{#1}}\fi
\ifx \doiurl  \undefined \def \doiurl#1{\url{https://doi.org/#1}}\fi
\ifx \betal  \undefined \def \betal{\textit{et al.}}\fi
\ifx \binstitute  \undefined \def \binstitute#1{#1}\fi
\ifx \binstitutionaled  \undefined \def \binstitutionaled#1{#1}\fi
\ifx \bctitle  \undefined \def \bctitle#1{#1}\fi
\ifx \beditor  \undefined \def \beditor#1{#1}\fi
\ifx \bpublisher  \undefined \def \bpublisher#1{#1}\fi
\ifx \bbtitle  \undefined \def \bbtitle#1{#1}\fi
\ifx \bedition  \undefined \def \bedition#1{#1}\fi
\ifx \bseriesno  \undefined \def \bseriesno#1{#1}\fi
\ifx \blocation  \undefined \def \blocation#1{#1}\fi
\ifx \bsertitle  \undefined \def \bsertitle#1{#1}\fi
\ifx \bsnm \undefined \def \bsnm#1{#1}\fi
\ifx \bsuffix \undefined \def \bsuffix#1{#1}\fi
\ifx \bparticle \undefined \def \bparticle#1{#1}\fi
\ifx \barticle \undefined \def \barticle#1{#1}\fi
\bibcommenthead
\ifx \bconfdate \undefined \def \bconfdate #1{#1}\fi
\ifx \botherref \undefined \def \botherref #1{#1}\fi
\ifx \url \undefined \def \url#1{\textsf{#1}}\fi
\ifx \bchapter \undefined \def \bchapter#1{#1}\fi
\ifx \bbook \undefined \def \bbook#1{#1}\fi
\ifx \bcomment \undefined \def \bcomment#1{#1}\fi
\ifx \oauthor \undefined \def \oauthor#1{#1}\fi
\ifx \citeauthoryear \undefined \def \citeauthoryear#1{#1}\fi
\ifx \endbibitem  \undefined \def \endbibitem {}\fi
\ifx \bconflocation  \undefined \def \bconflocation#1{#1}\fi
\ifx \arxivurl  \undefined \def \arxivurl#1{\textsf{#1}}\fi
\csname PreBibitemsHook\endcsname

\bibitem[\protect\citeauthoryear{{Serjeant}}{2023}]{Serjeant2023}
\begin{barticle}
\bauthor{\bsnm{{Serjeant}}, \binits{S.}}:
\batitle{{Citizen Science in the European Open Science Cloud}}.
\bjtitle{Europhysics News}
\bvolume{54}(\bissue{2}),
\bfpage{20}--\blpage{23}
(\byear{2023})
\doiurl{10.1051/epn/2023204}
{\href{https://arxiv.org/abs/2307.06896}{{arXiv:2307.06896}}}
{[astro-ph.IM]}
\end{barticle}
\endbibitem

\bibitem[\protect\citeauthoryear{{Swanson} et~al.}{2015}]{Swanson2015}
\begin{barticle}
\bauthor{\bsnm{{Swanson}}, \binits{A.}},
\bauthor{\bsnm{{Kosmala}}, \binits{M.}},
\bauthor{\bsnm{{Lintott}}, \binits{C.}},
\bauthor{\bsnm{{Simpson}}, \binits{R.}},
\bauthor{\bsnm{{Smith}}, \binits{A.}},
\bauthor{\bsnm{{Packer}}, \binits{C.}}:
\batitle{{Snapshot Serengeti, high-frequency annotated camera trap images of 40
  mammalian species in an African savanna}}.
\bjtitle{Scientific Data}
\bvolume{2},
\bfpage{150026}
(\byear{2015})
\doiurl{10.1038/sdata.2015.26}
\end{barticle}
\endbibitem

\bibitem[\protect\citeauthoryear{{Daylan} et~al.}{2016}]{Daylan2016}
\begin{barticle}
\bauthor{\bsnm{{Daylan}}, \binits{T.}},
\bauthor{\bsnm{{Finkbeiner}}, \binits{D.P.}},
\bauthor{\bsnm{{Hooper}}, \binits{D.}},
\bauthor{\bsnm{{Linden}}, \binits{T.}},
\bauthor{\bsnm{{Portillo}}, \binits{S.K.N.}},
\bauthor{\bsnm{{Rodd}}, \binits{N.L.}},
\bauthor{\bsnm{{Slatyer}}, \binits{T.R.}}:
\batitle{{The characterization of the gamma-ray signal from the central Milky
  Way: A case for annihilating dark matter}}.
\bjtitle{Physics of the Dark Universe}
\bvolume{12},
\bfpage{1}--\blpage{23}
(\byear{2016})
\doiurl{10.1016/j.dark.2015.12.005}
{\href{https://arxiv.org/abs/1402.6703}{{arXiv:1402.6703}}}
{[astro-ph.HE]}
\end{barticle}
\endbibitem

\bibitem[\protect\citeauthoryear{{Ackermann} et~al.}{2017}]{Ackermann2017}
\begin{barticle}
\bauthor{\bsnm{{Ackermann}}, \binits{M.}},
\bauthor{\bsnm{{Ajello}}, \binits{M.}},
\bauthor{\bsnm{{Albert}}, \binits{A.}},
\bauthor{\bsnm{{Atwood}}, \binits{W.B.}},
\bauthor{\bsnm{{Baldini}}, \binits{L.}},
\bauthor{\bsnm{{Ballet}}, \binits{J.}},
\bauthor{\bsnm{{Barbiellini}}, \binits{G.}},
\bauthor{\bsnm{{Bastieri}}, \binits{D.}},
\bauthor{\bsnm{{Bellazzini}}, \binits{R.}},
\bauthor{\bsnm{{Bissaldi}}, \binits{E.}},
\bauthor{\bsnm{{Blandford}}, \binits{R.D.}},
\bauthor{\bsnm{{Bloom}}, \binits{E.D.}},
\bauthor{\bsnm{{Bonino}}, \binits{R.}},
\bauthor{\bsnm{{Bottacini}}, \binits{E.}},
\bauthor{\bsnm{{Brandt}}, \binits{T.J.}},
\bauthor{\bsnm{{Bregeon}}, \binits{J.}},
\bauthor{\bsnm{{Bruel}}, \binits{P.}},
\bauthor{\bsnm{{Buehler}}, \binits{R.}},
\bauthor{\bsnm{{Burnett}}, \binits{T.H.}},
\bauthor{\bsnm{{Cameron}}, \binits{R.A.}},
\bauthor{\bsnm{{Caputo}}, \binits{R.}},
\bauthor{\bsnm{{Caragiulo}}, \binits{M.}},
\bauthor{\bsnm{{Caraveo}}, \binits{P.A.}},
\bauthor{\bsnm{{Cavazzuti}}, \binits{E.}},
\bauthor{\bsnm{{Cecchi}}, \binits{C.}},
\bauthor{\bsnm{{Charles}}, \binits{E.}},
\bauthor{\bsnm{{Chekhtman}}, \binits{A.}},
\bauthor{\bsnm{{Chiang}}, \binits{J.}},
\bauthor{\bsnm{{Chiappo}}, \binits{A.}},
\bauthor{\bsnm{{Chiaro}}, \binits{G.}},
\bauthor{\bsnm{{Ciprini}}, \binits{S.}},
\bauthor{\bsnm{{Conrad}}, \binits{J.}},
\bauthor{\bsnm{{Costanza}}, \binits{F.}},
\bauthor{\bsnm{{Cuoco}}, \binits{A.}},
\bauthor{\bsnm{{Cutini}}, \binits{S.}},
\bauthor{\bsnm{{D'Ammando}}, \binits{F.}},
\bauthor{\bsnm{{de Palma}}, \binits{F.}},
\bauthor{\bsnm{{Desiante}}, \binits{R.}},
\bauthor{\bsnm{{Digel}}, \binits{S.W.}},
\bauthor{\bsnm{{Di Lalla}}, \binits{N.}},
\bauthor{\bsnm{{Di Mauro}}, \binits{M.}},
\bauthor{\bsnm{{Di Venere}}, \binits{L.}},
\bauthor{\bsnm{{Drell}}, \binits{P.S.}},
\bauthor{\bsnm{{Favuzzi}}, \binits{C.}},
\bauthor{\bsnm{{Fegan}}, \binits{S.J.}},
\bauthor{\bsnm{{Ferrara}}, \binits{E.C.}},
\bauthor{\bsnm{{Focke}}, \binits{W.B.}},
\bauthor{\bsnm{{Franckowiak}}, \binits{A.}},
\bauthor{\bsnm{{Fukazawa}}, \binits{Y.}},
\bauthor{\bsnm{{Funk}}, \binits{S.}},
\bauthor{\bsnm{{Fusco}}, \binits{P.}},
\bauthor{\bsnm{{Gargano}}, \binits{F.}},
\bauthor{\bsnm{{Gasparrini}}, \binits{D.}},
\bauthor{\bsnm{{Giglietto}}, \binits{N.}},
\bauthor{\bsnm{{Giordano}}, \binits{F.}},
\bauthor{\bsnm{{Giroletti}}, \binits{M.}},
\bauthor{\bsnm{{Glanzman}}, \binits{T.}},
\bauthor{\bsnm{{Gomez-Vargas}}, \binits{G.A.}},
\bauthor{\bsnm{{Green}}, \binits{D.}},
\bauthor{\bsnm{{Grenier}}, \binits{I.A.}},
\bauthor{\bsnm{{Grove}}, \binits{J.E.}},
\bauthor{\bsnm{{Guillemot}}, \binits{L.}},
\bauthor{\bsnm{{Guiriec}}, \binits{S.}},
\bauthor{\bsnm{{Gustafsson}}, \binits{M.}},
\bauthor{\bsnm{{Harding}}, \binits{A.K.}},
\bauthor{\bsnm{{Hays}}, \binits{E.}},
\bauthor{\bsnm{{Hewitt}}, \binits{J.W.}},
\bauthor{\bsnm{{Horan}}, \binits{D.}},
\bauthor{\bsnm{{Jogler}}, \binits{T.}},
\bauthor{\bsnm{{Johnson}}, \binits{A.S.}},
\bauthor{\bsnm{{Kamae}}, \binits{T.}},
\bauthor{\bsnm{{Kocevski}}, \binits{D.}},
\bauthor{\bsnm{{Kuss}}, \binits{M.}},
\bauthor{\bsnm{{La Mura}}, \binits{G.}},
\bauthor{\bsnm{{Larsson}}, \binits{S.}},
\bauthor{\bsnm{{Latronico}}, \binits{L.}},
\bauthor{\bsnm{{Li}}, \binits{J.}},
\bauthor{\bsnm{{Longo}}, \binits{F.}},
\bauthor{\bsnm{{Loparco}}, \binits{F.}},
\bauthor{\bsnm{{Lovellette}}, \binits{M.N.}},
\bauthor{\bsnm{{Lubrano}}, \binits{P.}},
\bauthor{\bsnm{{Magill}}, \binits{J.D.}},
\bauthor{\bsnm{{Maldera}}, \binits{S.}},
\bauthor{\bsnm{{Malyshev}}, \binits{D.}},
\bauthor{\bsnm{{Manfreda}}, \binits{A.}},
\bauthor{\bsnm{{Martin}}, \binits{P.}},
\bauthor{\bsnm{{Mazziotta}}, \binits{M.N.}},
\bauthor{\bsnm{{Michelson}}, \binits{P.F.}},
\bauthor{\bsnm{{Mirabal}}, \binits{N.}},
\bauthor{\bsnm{{Mitthumsiri}}, \binits{W.}},
\bauthor{\bsnm{{Mizuno}}, \binits{T.}},
\bauthor{\bsnm{{Moiseev}}, \binits{A.A.}},
\bauthor{\bsnm{{Monzani}}, \binits{M.E.}},
\bauthor{\bsnm{{Morselli}}, \binits{A.}},
\bauthor{\bsnm{{Negro}}, \binits{M.}},
\bauthor{\bsnm{{Nuss}}, \binits{E.}},
\bauthor{\bsnm{{Ohsugi}}, \binits{T.}},
\bauthor{\bsnm{{Orienti}}, \binits{M.}},
\bauthor{\bsnm{{Orlando}}, \binits{E.}},
\bauthor{\bsnm{{Ormes}}, \binits{J.F.}},
\bauthor{\bsnm{{Paneque}}, \binits{D.}},
\bauthor{\bsnm{{Perkins}}, \binits{J.S.}},
\bauthor{\bsnm{{Persic}}, \binits{M.}},
\bauthor{\bsnm{{Pesce-Rollins}}, \binits{M.}},
\bauthor{\bsnm{{Piron}}, \binits{F.}},
\bauthor{\bsnm{{Principe}}, \binits{G.}},
\bauthor{\bsnm{{Rain{\`o}}}, \binits{S.}},
\bauthor{\bsnm{{Rando}}, \binits{R.}},
\bauthor{\bsnm{{Razzano}}, \binits{M.}},
\bauthor{\bsnm{{Razzaque}}, \binits{S.}},
\bauthor{\bsnm{{Reimer}}, \binits{A.}},
\bauthor{\bsnm{{Reimer}}, \binits{O.}},
\bauthor{\bsnm{{S{\'a}nchez-Conde}}, \binits{M.}},
\bauthor{\bsnm{{Sgr{\`o}}}, \binits{C.}},
\bauthor{\bsnm{{Simone}}, \binits{D.}},
\bauthor{\bsnm{{Siskind}}, \binits{E.J.}},
\bauthor{\bsnm{{Spada}}, \binits{F.}},
\bauthor{\bsnm{{Spandre}}, \binits{G.}},
\bauthor{\bsnm{{Spinelli}}, \binits{P.}},
\bauthor{\bsnm{{Suson}}, \binits{D.J.}},
\bauthor{\bsnm{{Tajima}}, \binits{H.}},
\bauthor{\bsnm{{Tanaka}}, \binits{K.}},
\bauthor{\bsnm{{Thayer}}, \binits{J.B.}},
\bauthor{\bsnm{{Tibaldo}}, \binits{L.}},
\bauthor{\bsnm{{Torres}}, \binits{D.F.}},
\bauthor{\bsnm{{Troja}}, \binits{E.}},
\bauthor{\bsnm{{Uchiyama}}, \binits{Y.}},
\bauthor{\bsnm{{Vianello}}, \binits{G.}},
\bauthor{\bsnm{{Wood}}, \binits{K.S.}},
\bauthor{\bsnm{{Wood}}, \binits{M.}},
\bauthor{\bsnm{{Zaharijas}}, \binits{G.}},
\bauthor{\bsnm{{Zimmer}}, \binits{S.}},
\bauthor{\bsnm{{Fermi LAT Collaboration}}}:
\batitle{{The Fermi Galactic Center GeV Excess and Implications for Dark
  Matter}}.
\bjtitle{\apj}
\bvolume{840}(\bissue{1}),
\bfpage{43}
(\byear{2017})
\doiurl{10.3847/1538-4357/aa6cab}
{\href{https://arxiv.org/abs/1704.03910}{{arXiv:1704.03910}}}
{[astro-ph.HE]}
\end{barticle}
\endbibitem

\bibitem[\protect\citeauthoryear{{Feng} et~al.}{2017}]{Feng2017}
\begin{bchapter}
\bauthor{\bsnm{{Feng}}, \binits{Q.}},
\bauthor{\bsnm{{Jarvis}}, \binits{J.}},
\bauthor{\bsnm{{VERITAS Collaboration}}}:
\bctitle{{A citizen-science approach to muon events in imaging atmospheric
  Cherenkov telescope data: the Muon Hunter}}.
In: \bbtitle{35th International Cosmic Ray Conference (ICRC2017)}.
\bsertitle{International Cosmic Ray Conference},
vol. \bseriesno{301},
p. \bfpage{826}
(\byear{2017}).
\doiurl{10.22323/1.301.0826}
\end{bchapter}
\endbibitem

\bibitem[\protect\citeauthoryear{{Bird} et~al.}{2018}]{Bird2018}
\begin{botherref}
\oauthor{\bsnm{{Bird}}, \binits{R.}},
\oauthor{\bsnm{{Daniel}}, \binits{M.K.}},
\oauthor{\bsnm{{Dickinson}}, \binits{H.}},
\oauthor{\bsnm{{Feng}}, \binits{Q.}},
\oauthor{\bsnm{{Fortson}}, \binits{L.}},
\oauthor{\bsnm{{Furniss}}, \binits{A.}},
\oauthor{\bsnm{{Jarvis}}, \binits{J.}},
\oauthor{\bsnm{{Mukherjee}}, \binits{R.}},
\oauthor{\bsnm{{Ong}}, \binits{R.}},
\oauthor{\bsnm{{Sadeh}}, \binits{I.}},
\oauthor{\bsnm{{Williams}}, \binits{D.}}:
{Muon Hunter: a Zooniverse project}.
arXiv e-prints,
1802--08907
(2018)
\doiurl{10.48550/arXiv.1802.08907}
{\href{https://arxiv.org/abs/1802.08907}{{arXiv:1802.08907}}}
{[astro-ph.IM]}
\end{botherref}
\endbibitem

\bibitem[\protect\citeauthoryear{{Laraia} et~al.}{2019}]{Laraia2019}
\begin{bchapter}
\bauthor{\bsnm{{Laraia}}, \binits{M.}},
\bauthor{\bsnm{{Wright}}, \binits{D.}},
\bauthor{\bsnm{{Dickinson}}, \binits{H.}},
\bauthor{\bsnm{{Simenstad}}, \binits{A.}},
\bauthor{\bsnm{{Flanagan}}, \binits{K.}},
\bauthor{\bsnm{{Serjeant}}, \binits{S.}},
\bauthor{\bsnm{{Fortson}}, \binits{L.}},
\bauthor{\bsnm{{VERITAS Collaboration}}}:
\bctitle{{Muon Hunter 2.0: efficient crowdsourcing of labels for IACT image
  analysis}}.
In: \bbtitle{36th International Cosmic Ray Conference (ICRC2019)}.
\bsertitle{International Cosmic Ray Conference},
vol. \bseriesno{36},
p. \bfpage{678}
(\byear{2019}).
\doiurl{10.22323/1.358.0678}
\end{bchapter}
\endbibitem

\bibitem[\protect\citeauthoryear{{Bird} et~al.}{2020}]{Bird2020}
\begin{bchapter}
\bauthor{\bsnm{{Bird}}, \binits{R.}},
\bauthor{\bsnm{{Daniel}}, \binits{M.K.}},
\bauthor{\bsnm{{Dickinson}}, \binits{H.}},
\bauthor{\bsnm{{Feng}}, \binits{Q.}},
\bauthor{\bsnm{{Fortson}}, \binits{L.}},
\bauthor{\bsnm{{Furniss}}, \binits{A.}},
\bauthor{\bsnm{{Jarvis}}, \binits{J.}},
\bauthor{\bsnm{{Mukherjee}}, \binits{R.}},
\bauthor{\bsnm{{Ong}}, \binits{R.}},
\bauthor{\bsnm{{Sadeh}}, \binits{I.}},
\bauthor{\bsnm{{Williams}}, \binits{D.}}:
\bctitle{{Muon Hunter: a Zooniverse project}}.
In: \bbtitle{Journal of Physics Conference Series}.
\bsertitle{Journal of Physics Conference Series},
vol. \bseriesno{1342},
p. \bfpage{012103}
(\byear{2020}).
\doiurl{10.1088/1742-6596/1342/1/012103}
\end{bchapter}
\endbibitem

\bibitem[\protect\citeauthoryear{{Flanagan} et~al.}{2022}]{Flanagan2022}
\begin{bchapter}
\bauthor{\bsnm{{Flanagan}}, \binits{K.}},
\bauthor{\bsnm{{Wright}}, \binits{D.}},
\bauthor{\bsnm{{Dickinson}}, \binits{H.}},
\bauthor{\bsnm{{Wilcox}}, \binits{P.}},
\bauthor{\bsnm{{Laraia}}, \binits{M.}},
\bauthor{\bsnm{{Serjeant}}, \binits{S.}},
\bauthor{\bsnm{{Capasso}}, \binits{M.}},
\bauthor{\bsnm{{Ong}}, \binits{R.}},
\bauthor{\bsnm{{Sadeh}}, \binits{I.}},
\bauthor{\bsnm{{Kaaret}}, \binits{P.}},
\bauthor{\bsnm{{Jin}}, \binits{W.}},
\bauthor{\bsnm{{Benbow}}, \binits{W.}},
\bauthor{\bsnm{{Mukherjee}}, \binits{R.}},
\bauthor{\bsnm{{Prado}}, \binits{R.}},
\bauthor{\bsnm{{Lundy}}, \binits{M.}},
\bauthor{\bsnm{{Patel}}, \binits{S.}},
\bauthor{\bsnm{{Moriarty}}, \binits{P.}},
\bauthor{\bsnm{{Maier}}, \binits{G.}},
\bauthor{\bsnm{{Furniss}}, \binits{A.}},
\bauthor{\bsnm{{Ragan}}, \binits{K.}},
\bauthor{\bsnm{{Williams}}, \binits{D.}},
\bauthor{\bsnm{{Buckley}}, \binits{J.}},
\bauthor{\bsnm{{Fortson}}, \binits{L.}},
\bauthor{\bsnm{{Quinn}}, \binits{J.}},
\bauthor{\bsnm{{Holder}}, \binits{J.}},
\bauthor{\bsnm{{Giuri}}, \binits{C.}},
\bauthor{\bsnm{{Pueschel}}, \binits{E.}},
\bauthor{\bsnm{{Nieto}}, \binits{D.}},
\bauthor{\bsnm{{Adams}}, \binits{C.}},
\bauthor{\bsnm{{O'Brien}}, \binits{S.}},
\bauthor{\bsnm{{Ribeiro}}, \binits{D.}},
\bauthor{\bsnm{{Pfrang}}, \binits{K.}},
\bauthor{\bsnm{{Gueta}}, \binits{O.}},
\bauthor{\bsnm{{Foote}}, \binits{G.}},
\bauthor{\bsnm{{Weinstein}}, \binits{A.}},
\bauthor{\bsnm{{Kumar}}, \binits{S.}},
\bauthor{\bsnm{{Williamson}}, \binits{T.}},
\bauthor{\bsnm{{Tak}}, \binits{D.}},
\bauthor{\bsnm{{McGrath}}, \binits{C.}},
\bauthor{\bsnm{{Kleiner}}, \binits{T.}},
\bauthor{\bsnm{{Pohl}}, \binits{M.}},
\bauthor{\bsnm{{Reynolds}}, \binits{P.}},
\bauthor{\bsnm{{Hona}}, \binits{B.}},
\bauthor{\bsnm{{Hanna}}, \binits{D.}},
\bauthor{\bsnm{{Santander}}, \binits{M.}},
\bauthor{\bsnm{{Sembroski}}, \binits{G.}},
\bauthor{\bsnm{{Patel}}, \binits{S.R.}},
\bauthor{\bsnm{{Errando}}, \binits{M.}},
\bauthor{\bsnm{{Kertzman}}, \binits{M.}},
\bauthor{\bsnm{{Hervet}}, \binits{O.}},
\bauthor{\bsnm{{Nievas-Rosillo}}, \binits{M.}},
\bauthor{\bsnm{{Lang}}, \binits{M.}},
\bauthor{\bsnm{{Roache}}, \binits{E.}},
\bauthor{\bsnm{{Humensky}}, \binits{T.B.}},
\bauthor{\bsnm{{Shang}}, \binits{R.Y.}},
\bauthor{\bsnm{{Vassiliev}}, \binits{V.}},
\bauthor{\bsnm{{Chromey}}, \binits{A.}},
\bauthor{\bsnm{{Falcone}}, \binits{A.}},
\bauthor{\bsnm{{Christiansen}}, \binits{J.}},
\bauthor{\bsnm{{Otte}}, \binits{A.}},
\bauthor{\bsnm{{Gent}}, \binits{A.E.}},
\bauthor{\bsnm{{Brill}}, \binits{A.}},
\bauthor{\bsnm{{Ryan}}, \binits{J.}},
\bauthor{\bsnm{{Farrell}}, \binits{K.}},
\bauthor{\bsnm{{Gillanders}}, \binits{G.}},
\bauthor{\bsnm{{Feng}}, \binits{Q.}},
\bauthor{\bsnm{{Archer}}, \binits{A.}},
\bauthor{\bsnm{{Kieda}}, \binits{D.}}:
\bctitle{{Identifying muon rings in VERITAS data using convolutional neural
  networks trained on images classified with Muon Hunters 2}}.
In: \bbtitle{37th International Cosmic Ray Conference},
p. \bfpage{766}
(\byear{2022}).
\doiurl{10.22323/1.395.0766}
\end{bchapter}
\endbibitem

\bibitem[\protect\citeauthoryear{{Norton}}{2018}]{Norton2018}
\begin{barticle}
\bauthor{\bsnm{{Norton}}, \binits{A.J.}}:
\batitle{{A Zooniverse Project to Classify Periodic Variable Stars from
  SuperWASP}}.
\bjtitle{Research Notes of the American Astronomical Society}
\bvolume{2}(\bissue{4}),
\bfpage{216}
(\byear{2018})
\doiurl{10.3847/2515-5172/aaf291}
\end{barticle}
\endbibitem

\bibitem[\protect\citeauthoryear{{Thiemann} et~al.}{2021}]{Thiemann2021}
\begin{barticle}
\bauthor{\bsnm{{Thiemann}}, \binits{H.B.}},
\bauthor{\bsnm{{Norton}}, \binits{A.J.}},
\bauthor{\bsnm{{Dickinson}}, \binits{H.J.}},
\bauthor{\bsnm{{McMaster}}, \binits{A.}},
\bauthor{\bsnm{{Kolb}}, \binits{U.C.}}:
\batitle{{SuperWASP variable stars: classifying light curves using citizen
  science}}.
\bjtitle{\mnras}
\bvolume{502}(\bissue{1}),
\bfpage{1299}--\blpage{1311}
(\byear{2021})
\doiurl{10.1093/mnras/stab140}
{\href{https://arxiv.org/abs/2101.06216}{{arXiv:2101.06216}}}
{[astro-ph.SR]}
\end{barticle}
\endbibitem

\bibitem[\protect\citeauthoryear{{Wo{\'z}niak}}{2019}]{credo2019}
\begin{bchapter}
\bauthor{\bsnm{{Wo{\'z}niak}}, \binits{K.W.}}:
\bctitle{{Detection of Cosmic-Ray Ensembles with CREDO}}.
In: \bbtitle{European Physical Journal Web of Conferences}.
\bsertitle{European Physical Journal Web of Conferences},
vol. \bseriesno{208},
p. \bfpage{15006}
(\byear{2019}).
\doiurl{10.1051/epjconf/201920815006}
\end{bchapter}
\endbibitem

\bibitem[\protect\citeauthoryear{{Wibig} et~al.}{2022}]{credo2022}
\begin{bchapter}
\bauthor{\bsnm{{Wibig}}, \binits{T.}},
\bauthor{\bsnm{{Karbowiak}}, \binits{M.}},
\bauthor{\bsnm{{Alvarez-Castillo}}, \binits{D.}},
\bauthor{\bsnm{{Bar}}, \binits{O.}},
\bauthor{\bsnm{{Bibrzycki}}, \binits{{\L}.}},
\bauthor{\bsnm{{Gora}}, \binits{D.}},
\bauthor{\bsnm{{Homola}}, \binits{P.}},
\bauthor{\bsnm{{Kovacs}}, \binits{P.}},
\bauthor{\bsnm{{Piekarczyk}}, \binits{M.}},
\bauthor{\bsnm{{Stasielak}}, \binits{J.}},
\bauthor{\bsnm{{Stuglik}}, \binits{S.}},
\bauthor{\bsnm{{Sushchov}}, \binits{O.}},
\bauthor{\bsnm{{Tursunov}}, \binits{A.}}:
\bctitle{{Determination of Zenith Angle Dependence of Incoherent Cosmic Ray
  Muon Flux Using Smartphones of the CREDO Project}}.
In: \bbtitle{37th International Cosmic Ray Conference},
p. \bfpage{199}
(\byear{2022}).
\doiurl{10.22323/1.395.0199}
\end{bchapter}
\endbibitem

\bibitem[\protect\citeauthoryear{{Tursunov} et~al.}{2022}]{credo2022b}
\begin{bchapter}
\bauthor{\bsnm{{Tursunov}}, \binits{A.}},
\bauthor{\bsnm{{Homola}}, \binits{P.}},
\bauthor{\bsnm{{Alvarez-Castillo}}, \binits{D.}},
\bauthor{\bsnm{{Budnev}}, \binits{N.}},
\bauthor{\bsnm{{Gupta}}, \binits{A.}},
\bauthor{\bsnm{{Hnatyk}}, \binits{B.}},
\bauthor{\bsnm{{Kasztelan}}, \binits{M.}},
\bauthor{\bsnm{{Kovacs}}, \binits{P.}},
\bauthor{\bsnm{{{\L}ozowski}}, \binits{B.}},
\bauthor{\bsnm{{Medvedev}}, \binits{M.}},
\bauthor{\bsnm{{Mozgova}}, \binits{A.}},
\bauthor{\bsnm{{Niedzwiecki}}, \binits{M.}},
\bauthor{\bsnm{{Pawlik}}, \binits{M.}},
\bauthor{\bsnm{{Rosas}}, \binits{M.}},
\bauthor{\bsnm{{Rzecki}}, \binits{K.}},
\bauthor{\bsnm{{Smelcerz}}, \binits{K.}},
\bauthor{\bsnm{{Smolek}}, \binits{K.}},
\bauthor{\bsnm{{Stasielak}}, \binits{J.}},
\bauthor{\bsnm{{Stuglik}}, \binits{S.}},
\bauthor{\bsnm{{Svanidze}}, \binits{M.}},
\bauthor{\bsnm{{Sushchov}}, \binits{O.}},
\bauthor{\bsnm{{Verbetsky}}, \binits{Y.}},
\bauthor{\bsnm{{Wibig}}, \binits{T.}},
\bauthor{\bsnm{{Zamora-Saa.}}, \binits{J.}},
\bauthor{\bsnm{{Credo Collaboration}}}:
\bctitle{{Probing UHECR and cosmic ray ensemble scenarios with a global CREDO
  network}}.
In: \bbtitle{37th International Cosmic Ray Conference},
p. \bfpage{472}
(\byear{2022}).
\doiurl{10.22323/1.395.0472}
\end{bchapter}
\endbibitem

\bibitem[\protect\citeauthoryear{{Metcalf} et~al.}{2019}]{Metcalf+19}
\begin{barticle}
\bauthor{\bsnm{{Metcalf}}, \binits{R.B.}},
\bauthor{\bsnm{{Meneghetti}}, \binits{M.}},
\bauthor{\bsnm{{Avestruz}}, \binits{C.}},
\bauthor{\bsnm{{Bellagamba}}, \binits{F.}},
\bauthor{\bsnm{{Bom}}, \binits{C.R.}},
\bauthor{\bsnm{{Bertin}}, \binits{E.}},
\bauthor{\bsnm{{Cabanac}}, \binits{R.}},
\bauthor{\bsnm{{Courbin}}, \binits{F.}},
\bauthor{\bsnm{{Davies}}, \binits{A.}},
\bauthor{\bsnm{{Decenci{\`e}re}}, \binits{E.}},
\bauthor{\bsnm{{Flamary}}, \binits{R.}},
\bauthor{\bsnm{{Gavazzi}}, \binits{R.}},
\bauthor{\bsnm{{Geiger}}, \binits{M.}},
\bauthor{\bsnm{{Hartley}}, \binits{P.}},
\bauthor{\bsnm{{Huertas-Company}}, \binits{M.}},
\bauthor{\bsnm{{Jackson}}, \binits{N.}},
\bauthor{\bsnm{{Jacobs}}, \binits{C.}},
\bauthor{\bsnm{{Jullo}}, \binits{E.}},
\bauthor{\bsnm{{Kneib}}, \binits{J.-P.}},
\bauthor{\bsnm{{Koopmans}}, \binits{L.V.E.}},
\bauthor{\bsnm{{Lanusse}}, \binits{F.}},
\bauthor{\bsnm{{Li}}, \binits{C.-L.}},
\bauthor{\bsnm{{Ma}}, \binits{Q.}},
\bauthor{\bsnm{{Makler}}, \binits{M.}},
\bauthor{\bsnm{{Li}}, \binits{N.}},
\bauthor{\bsnm{{Lightman}}, \binits{M.}},
\bauthor{\bsnm{{Petrillo}}, \binits{C.E.}},
\bauthor{\bsnm{{Serjeant}}, \binits{S.}},
\bauthor{\bsnm{{Sch{\"a}fer}}, \binits{C.}},
\bauthor{\bsnm{{Sonnenfeld}}, \binits{A.}},
\bauthor{\bsnm{{Tagore}}, \binits{A.}},
\bauthor{\bsnm{{Tortora}}, \binits{C.}},
\bauthor{\bsnm{{Tuccillo}}, \binits{D.}},
\bauthor{\bsnm{{Valent{\'\i}n}}, \binits{M.B.}},
\bauthor{\bsnm{{Velasco-Forero}}, \binits{S.}},
\bauthor{\bsnm{{Verdoes Kleijn}}, \binits{G.A.}},
\bauthor{\bsnm{{Vernardos}}, \binits{G.}}:
\batitle{{The strong gravitational lens finding challenge}}.
\bjtitle{\aap}
\bvolume{625},
\bfpage{119}
(\byear{2019})
\doiurl{10.1051/0004-6361/201832797}
{\href{https://arxiv.org/abs/1802.03609}{{arXiv:1802.03609}}}
{[astro-ph.GA]}
\end{barticle}
\endbibitem

\bibitem[\protect\citeauthoryear{Davies}{2022}]{DaviesPhD}
\begin{botherref}
\oauthor{\bsnm{Davies}, \binits{A.}}:
Using machine learning techniques to detect, classify and separate strong
  gravitational lensing systems from astronomical images.
PhD thesis,
The Open University,
The Open University, Milton Keynes, MK7 6AA, UK
(2022)
\end{botherref}
\endbibitem

\bibitem[\protect\citeauthoryear{{Dickinson} et~al.}{2022}]{Dickinson2022}
\begin{barticle}
\bauthor{\bsnm{{Dickinson}}, \binits{H.}},
\bauthor{\bsnm{{Adams}}, \binits{D.}},
\bauthor{\bsnm{{Mehta}}, \binits{V.}},
\bauthor{\bsnm{{Scarlata}}, \binits{C.}},
\bauthor{\bsnm{{Fortson}}, \binits{L.}},
\bauthor{\bsnm{{Serjeant}}, \binits{S.}},
\bauthor{\bsnm{{Krawczyk}}, \binits{C.}},
\bauthor{\bsnm{{Kruk}}, \binits{S.}},
\bauthor{\bsnm{{Lintott}}, \binits{C.}},
\bauthor{\bsnm{{Mantha}}, \binits{K.B.}},
\bauthor{\bsnm{{Simmons}}, \binits{B.D.}},
\bauthor{\bsnm{{Walmsley}}, \binits{M.}}:
\batitle{{Galaxy Zoo: Clump Scout - Design and first application of a
  two-dimensional aggregation tool for citizen science}}.
\bjtitle{\mnras}
\bvolume{517}(\bissue{4}),
\bfpage{5882}--\blpage{5911}
(\byear{2022})
\doiurl{10.1093/mnras/stac2919}
{\href{https://arxiv.org/abs/2210.03684}{{arXiv:2210.03684}}}
{[astro-ph.GA]}
\end{barticle}
\endbibitem

\bibitem[\protect\citeauthoryear{{Adams} et~al.}{2022}]{Adams2022}
\begin{barticle}
\bauthor{\bsnm{{Adams}}, \binits{D.}},
\bauthor{\bsnm{{Mehta}}, \binits{V.}},
\bauthor{\bsnm{{Dickinson}}, \binits{H.}},
\bauthor{\bsnm{{Scarlata}}, \binits{C.}},
\bauthor{\bsnm{{Fortson}}, \binits{L.}},
\bauthor{\bsnm{{Kruk}}, \binits{S.}},
\bauthor{\bsnm{{Simmons}}, \binits{B.}},
\bauthor{\bsnm{{Lintott}}, \binits{C.}}:
\batitle{{Galaxy Zoo: Clump Scout: Surveying the Local Universe for Giant
  Star-forming Clumps}}.
\bjtitle{\apj}
\bvolume{931}(\bissue{1}),
\bfpage{16}
(\byear{2022})
\doiurl{10.3847/1538-4357/ac6512}
{\href{https://arxiv.org/abs/2201.06581}{{arXiv:2201.06581}}}
{[astro-ph.GA]}
\end{barticle}
\endbibitem

\bibitem[\protect\citeauthoryear{Prather et~al.}{2013}]{Prather2013}
\begin{botherref}
\oauthor{\bsnm{Prather}, \binits{E.E.}},
\oauthor{\bsnm{Cormier}, \binits{S.}},
\oauthor{\bsnm{Wallace}, \binits{C.S.}},
\oauthor{\bsnm{Lintott}, \binits{C.}},
\oauthor{\bsnm{Jordan~Raddick}, \binits{M.}},
\oauthor{\bsnm{Smith}, \binits{A.}}:
Measuring the conceptual understandings of citizen scientists participating in
  zooniverse projects: A first approach.
Astronomy Education Review
\textbf{12}(1)
(2013)
\doiurl{10.3847/aer2013002}
\end{botherref}
\endbibitem

\bibitem[\protect\citeauthoryear{Luczak-Roesch et~al.}{2014}]{LuczakRoesch2014}
\begin{barticle}
\bauthor{\bsnm{Luczak-Roesch}, \binits{M.}},
\bauthor{\bsnm{Tinati}, \binits{R.}},
\bauthor{\bsnm{Simperl}, \binits{E.}},
\bauthor{\bsnm{Van~Kleek}, \binits{M.}},
\bauthor{\bsnm{Shadbolt}, \binits{N.}},
\bauthor{\bsnm{Simpson}, \binits{R.}}:
\batitle{Why won’t aliens talk to us? content and community dynamics in
  online citizen science}.
\bjtitle{Proceedings of the International AAAI Conference on Web and Social
  Media}
\bvolume{8}(\bissue{1}),
\bfpage{315}--\blpage{324}
(\byear{2014})
\doiurl{10.1609/icwsm.v8i1.14539}
\end{barticle}
\endbibitem

\bibitem[\protect\citeauthoryear{Masters et~al.}{2016}]{Masters+16}
\begin{botherref}
\oauthor{\bsnm{Masters}, \binits{K.}},
\oauthor{\bsnm{Oh}, \binits{E.Y.}},
\oauthor{\bsnm{Cox}, \binits{J.}},
\oauthor{\bsnm{Simmons}, \binits{B.}},
\oauthor{\bsnm{Lintott}, \binits{C.}},
\oauthor{\bsnm{Graham}, \binits{G.}},
\oauthor{\bsnm{Greenhill}, \binits{A.}},
\oauthor{\bsnm{Holmes}, \binits{K.}}:
Science Learning via Participation in Online Citizen Science.
arXiv
(2016).
\doiurl{10.48550/ARXIV.1601.05973} .
\url{https://arxiv.org/abs/1601.05973}
\end{botherref}
\endbibitem

\bibitem[\protect\citeauthoryear{Trouille et~al.}{2019}]{Trouille+19}
\begin{bchapter}
\bauthor{\bsnm{Trouille}, \binits{L.}},
\bauthor{\bsnm{Nelson}, \binits{T.}},
\bauthor{\bsnm{Feldt}, \binits{J.}},
\bauthor{\bsnm{Keller}, \binits{J.}},
\bauthor{\bsnm{Buie}, \binits{M.}},
\bauthor{\bsnm{Cardamone}, \binits{C.}},
\bauthor{\bsnm{Kung}, \binits{B.C.}},
\bauthor{\bsnm{Masters}, \binits{K.}},
\bauthor{\bsnm{Meredith}, \binits{K.}},
\bauthor{\bsnm{Borden}, \binits{K.}}:
\bctitle{Citizen science in astronomy education}.
In: \bbtitle{Astronomy Education, Volume 1}.
\bsertitle{2514-3433},
pp. \bfpage{8}--\blpage{1824}.
\bpublisher{IOP Publishing}, \blocation{???}
(\byear{2019}).
\doiurl{10.1088/2514-3433/ab2b42ch8} .
\burl{https://dx.doi.org/10.1088/2514-3433/ab2b42ch8}
\end{bchapter}
\endbibitem

\bibitem[\protect\citeauthoryear{Senabre~Hidalgo
  et~al.}{2021}]{SenabreHidalgo2021}
\begin{bbook}
\bauthor{\bsnm{Senabre~Hidalgo}, \binits{E.}},
\bauthor{\bsnm{Perell{\'o}}, \binits{J.}},
\bauthor{\bsnm{Becker}, \binits{F.}},
\bauthor{\bsnm{Bonhoure}, \binits{I.}},
\bauthor{\bsnm{Legris}, \binits{M.}},
\bauthor{\bsnm{Cigarini}, \binits{A.}}:
In: \beditor{\bsnm{Vohland}, \binits{K.}},
\beditor{\bsnm{Land-Zandstra}, \binits{A.}},
\beditor{\bsnm{Ceccaroni}, \binits{L.}},
\beditor{\bsnm{Lemmens}, \binits{R.}},
\beditor{\bsnm{Perell{\'o}}, \binits{J.}},
\beditor{\bsnm{Ponti}, \binits{M.}},
\beditor{\bsnm{Samson}, \binits{R.}},
\beditor{\bsnm{Wagenknecht}, \binits{K.}} (eds.)
\bbtitle{Participation and Co-creation in Citizen Science},
pp. \bfpage{199}--\blpage{218}.
\bpublisher{Springer},
\blocation{Cham}
(\byear{2021}).
\doiurl{10.1007/978-3-030-58278-4_11} .
\burl{https://doi.org/10.1007/978-3-030-58278-4_11}
\end{bbook}
\endbibitem

\bibitem[\protect\citeauthoryear{Keel et~al.}{2022}]{Keel+22}
\begin{barticle}
\bauthor{\bsnm{Keel}, \binits{W.C.}},
\bauthor{\bsnm{Tate}, \binits{J.}},
\bauthor{\bsnm{Wong}, \binits{O.I.}},
\bauthor{\bsnm{Banfield}, \binits{J.K.}},
\bauthor{\bsnm{Lintott}, \binits{C.J.}},
\bauthor{\bsnm{Masters}, \binits{K.L.}},
\bauthor{\bsnm{Simmons}, \binits{B.D.}},
\bauthor{\bsnm{Scarlata}, \binits{C.}},
\bauthor{\bsnm{Cardamone}, \binits{C.}},
\bauthor{\bsnm{Smethurst}, \binits{R.}},
\bauthor{\bsnm{Fortson}, \binits{L.}},
\bauthor{\bsnm{Shanahan}, \binits{J.}},
\bauthor{\bsnm{Kruk}, \binits{S.}},
\bauthor{\bsnm{Garland}, \binits{I.L.}},
\bauthor{\bsnm{Hancock}, \binits{C.}},
\bauthor{\bsnm{O’Ryan}, \binits{D.}}:
\batitle{Gems of the galaxy zoos—a wide-ranging hubble space telescope
  gap-filler program*}.
\bjtitle{The Astronomical Journal}
\bvolume{163}(\bissue{4}),
\bfpage{150}
(\byear{2022})
\doiurl{10.3847/1538-3881/ac517d}
\end{barticle}
\endbibitem

\bibitem[\protect\citeauthoryear{{Lintott} et~al.}{2009}]{Lintott+09}
\begin{barticle}
\bauthor{\bsnm{{Lintott}}, \binits{C.J.}},
\bauthor{\bsnm{{Schawinski}}, \binits{K.}},
\bauthor{\bsnm{{Keel}}, \binits{W.}},
\bauthor{\bsnm{{van Arkel}}, \binits{H.}},
\bauthor{\bsnm{{Bennert}}, \binits{N.}},
\bauthor{\bsnm{{Edmondson}}, \binits{E.}},
\bauthor{\bsnm{{Thomas}}, \binits{D.}},
\bauthor{\bsnm{{Smith}}, \binits{D.J.B.}},
\bauthor{\bsnm{{Herbert}}, \binits{P.D.}},
\bauthor{\bsnm{{Jarvis}}, \binits{M.J.}},
\bauthor{\bsnm{{Virani}}, \binits{S.}},
\bauthor{\bsnm{{Andreescu}}, \binits{D.}},
\bauthor{\bsnm{{Bamford}}, \binits{S.P.}},
\bauthor{\bsnm{{Land}}, \binits{K.}},
\bauthor{\bsnm{{Murray}}, \binits{P.}},
\bauthor{\bsnm{{Nichol}}, \binits{R.C.}},
\bauthor{\bsnm{{Raddick}}, \binits{M.J.}},
\bauthor{\bsnm{{Slosar}}, \binits{A.}},
\bauthor{\bsnm{{Szalay}}, \binits{A.}},
\bauthor{\bsnm{{Vandenberg}}, \binits{J.}}:
\batitle{{Galaxy Zoo: `Hanny's Voorwerp', a quasar light echo?}}
\bjtitle{\mnras}
\bvolume{399}(\bissue{1}),
\bfpage{129}--\blpage{140}
(\byear{2009})
\doiurl{10.1111/j.1365-2966.2009.15299.x}
{\href{https://arxiv.org/abs/0906.5304}{{arXiv:0906.5304}}}
{[astro-ph.CO]}
\end{barticle}
\endbibitem

\bibitem[\protect\citeauthoryear{{Cardamone} et~al.}{2009}]{Cardamone+09}
\begin{barticle}
\bauthor{\bsnm{{Cardamone}}, \binits{C.}},
\bauthor{\bsnm{{Schawinski}}, \binits{K.}},
\bauthor{\bsnm{{Sarzi}}, \binits{M.}},
\bauthor{\bsnm{{Bamford}}, \binits{S.P.}},
\bauthor{\bsnm{{Bennert}}, \binits{N.}},
\bauthor{\bsnm{{Urry}}, \binits{C.M.}},
\bauthor{\bsnm{{Lintott}}, \binits{C.}},
\bauthor{\bsnm{{Keel}}, \binits{W.C.}},
\bauthor{\bsnm{{Parejko}}, \binits{J.}},
\bauthor{\bsnm{{Nichol}}, \binits{R.C.}},
\bauthor{\bsnm{{Thomas}}, \binits{D.}},
\bauthor{\bsnm{{Andreescu}}, \binits{D.}},
\bauthor{\bsnm{{Murray}}, \binits{P.}},
\bauthor{\bsnm{{Raddick}}, \binits{M.J.}},
\bauthor{\bsnm{{Slosar}}, \binits{A.}},
\bauthor{\bsnm{{Szalay}}, \binits{A.}},
\bauthor{\bsnm{{Vandenberg}}, \binits{J.}}:
\batitle{{Galaxy Zoo Green Peas: discovery of a class of compact extremely
  star-forming galaxies}}.
\bjtitle{\mnras}
\bvolume{399}(\bissue{3}),
\bfpage{1191}--\blpage{1205}
(\byear{2009})
\doiurl{10.1111/j.1365-2966.2009.15383.x}
{\href{https://arxiv.org/abs/0907.4155}{{arXiv:0907.4155}}}
{[astro-ph.CO]}
\end{barticle}
\endbibitem

\bibitem[\protect\citeauthoryear{{De La Vega} et~al.}{2023}]{SonoUno}
\begin{botherref}
\oauthor{\bsnm{{De La Vega}}, \binits{G.}},
\oauthor{\bsnm{{Exequiel Dominguez}}, \binits{L.M.}},
\oauthor{\bsnm{{Casado}}, \binits{J.}},
\oauthor{\bsnm{{Garc{\'\i}a}}, \binits{B.}}:
{SonoUno web: an innovative user centred web interface}.
arXiv e-prints,
2302--00081
(2023)
\doiurl{10.48550/arXiv.2302.00081}
{\href{https://arxiv.org/abs/2302.00081}{{arXiv:2302.00081}}}
{[astro-ph.IM]}
\end{botherref}
\endbibitem

\bibitem[\protect\citeauthoryear{{Zanella} et~al.}{2022}]{Zanella+22}
\begin{barticle}
\bauthor{\bsnm{{Zanella}}, \binits{A.}},
\bauthor{\bsnm{{Harrison}}, \binits{C.M.}},
\bauthor{\bsnm{{Lenzi}}, \binits{S.}},
\bauthor{\bsnm{{Cooke}}, \binits{J.}},
\bauthor{\bsnm{{Damsma}}, \binits{P.}},
\bauthor{\bsnm{{Fleming}}, \binits{S.W.}}:
\batitle{{Sonification and sound design for astronomy research, education and
  public engagement}}.
\bjtitle{Nature Astronomy}
\bvolume{6},
\bfpage{1241}--\blpage{1248}
(\byear{2022})
\doiurl{10.1038/s41550-022-01721-z}
{\href{https://arxiv.org/abs/2206.13536}{{arXiv:2206.13536}}}
{[astro-ph.IM]}
\end{barticle}
\endbibitem

\bibitem[\protect\citeauthoryear{semenzin et~al.}{2020a}]{babysounds1}
\begin{botherref}
\oauthor{\bsnm{semenzin}, \binits{c.}},
\oauthor{\bsnm{Hamrick}, \binits{L.}},
\oauthor{\bsnm{Seidl}, \binits{A.}},
\oauthor{\bsnm{Kelleher}, \binits{B.L.}},
\oauthor{\bsnm{Cristia}, \binits{A.}}:
Describing vocalizations in young children: A big data approach through citizen
  science annotation.
OSF Preprints
(2020).
\doiurl{10.31219/osf.io/z6exv} .
\url{osf.io/z6exv}
\end{botherref}
\endbibitem

\bibitem[\protect\citeauthoryear{semenzin et~al.}{2020b}]{babysounds2}
\begin{botherref}
\oauthor{\bsnm{semenzin}, \binits{c.}},
\oauthor{\bsnm{Hamrick}, \binits{L.}},
\oauthor{\bsnm{Seidl}, \binits{A.}},
\oauthor{\bsnm{Kelleher}, \binits{B.L.}},
\oauthor{\bsnm{Cristia}, \binits{A.}}:
Towards Large-Scale Data Annotation of Audio from Wearables: Validating
  Zooniverse Annotations of Infant Vocalization Types.
OSF Preprints
(2020).
\doiurl{10.31219/osf.io/gpxf5} .
\url{osf.io/gpxf5}
\end{botherref}
\endbibitem

\bibitem[\protect\citeauthoryear{Cartwright et~al.}{2019}]{SoundsOfNewYork}
\begin{bchapter}
\bauthor{\bsnm{Cartwright}, \binits{M.}},
\bauthor{\bsnm{Dove}, \binits{G.}},
\bauthor{\bsnm{M\'{e}ndez~M\'{e}ndez}, \binits{A.E.}},
\bauthor{\bsnm{Bello}, \binits{J.P.}},
\bauthor{\bsnm{Nov}, \binits{O.}}:
\bctitle{Crowdsourcing multi-label audio annotation tasks with citizen
  scientists}.
In: \bbtitle{Proceedings of the 2019 CHI Conference on Human Factors in
  Computing Systems}.
\bsertitle{CHI '19},
pp. \bfpage{1}--\blpage{11}.
\bpublisher{Association for Computing Machinery},
\blocation{New York, NY, USA}
(\byear{2019}).
\doiurl{10.1145/3290605.3300522} .
\burl{https://doi.org/10.1145/3290605.3300522}
\end{bchapter}
\endbibitem

\bibitem[\protect\citeauthoryear{Chapman}{2017}]{AllanChapman}
\begin{bbook}
\bauthor{\bsnm{Chapman}, \binits{A.}}:
\bbtitle{Victorian Amateur Astronomer: Independent Astronomical Research in
  Britain 1820 - 1920},
(\byear{2017})
\end{bbook}
\endbibitem

\bibitem[\protect\citeauthoryear{{Rendtel}}{2017}]{Rendtel17}
\begin{barticle}
\bauthor{\bsnm{{Rendtel}}, \binits{J.}}:
\batitle{{Review of amateur meteor research}}.
\bjtitle{\planss}
\bvolume{143},
\bfpage{7}--\blpage{11}
(\byear{2017})
\doiurl{10.1016/j.pss.2017.01.007}
\end{barticle}
\endbibitem

\bibitem[\protect\citeauthoryear{{King} et~al.}{2022}]{King+22}
\begin{barticle}
\bauthor{\bsnm{{King}}, \binits{A.J.}},
\bauthor{\bsnm{{Daly}}, \binits{L.}},
\bauthor{\bsnm{{Rowe}}, \binits{J.}},
\bauthor{\bsnm{{Joy}}, \binits{K.H.}},
\bauthor{\bsnm{{Greenwood}}, \binits{R.C.}},
\bauthor{\bsnm{{Devillepoix}}, \binits{H.A.R.}},
\bauthor{\bsnm{{Suttle}}, \binits{M.D.}},
\bauthor{\bsnm{{Chan}}, \binits{Q.H.S.}},
\bauthor{\bsnm{{Russell}}, \binits{S.S.}},
\bauthor{\bsnm{{Bates}}, \binits{H.C.}},
\bauthor{\bsnm{{Bryson}}, \binits{J.F.J.}},
\bauthor{\bsnm{{Clay}}, \binits{P.L.}},
\bauthor{\bsnm{{Vida}}, \binits{D.}},
\bauthor{\bsnm{{Lee}}, \binits{M.R.}},
\bauthor{\bsnm{{O'Brien}}, \binits{{\'A}.}},
\bauthor{\bsnm{{Hallis}}, \binits{L.J.}},
\bauthor{\bsnm{{Stephen}}, \binits{N.R.}},
\bauthor{\bsnm{{Tart{\`e}se}}, \binits{R.}},
\bauthor{\bsnm{{Sansom}}, \binits{E.K.}},
\bauthor{\bsnm{{Towner}}, \binits{M.C.}},
\bauthor{\bsnm{{Cupak}}, \binits{M.}},
\bauthor{\bsnm{{Shober}}, \binits{P.M.}},
\bauthor{\bsnm{{Bland}}, \binits{P.A.}},
\bauthor{\bsnm{{Findlay}}, \binits{R.}},
\bauthor{\bsnm{{Franchi}}, \binits{I.A.}},
\bauthor{\bsnm{{Verchovsky}}, \binits{A.B.}},
\bauthor{\bsnm{{Abernethy}}, \binits{F.A.J.}},
\bauthor{\bsnm{{Grady}}, \binits{M.M.}},
\bauthor{\bsnm{{Floyd}}, \binits{C.J.}},
\bauthor{\bsnm{{Van Ginneken}}, \binits{M.}},
\bauthor{\bsnm{{Bridges}}, \binits{J.}},
\bauthor{\bsnm{{Hicks}}, \binits{L.J.}},
\bauthor{\bsnm{{Jones}}, \binits{R.H.}},
\bauthor{\bsnm{{Mitchell}}, \binits{J.T.}},
\bauthor{\bsnm{{Genge}}, \binits{M.J.}},
\bauthor{\bsnm{{Jenkins}}, \binits{L.}},
\bauthor{\bsnm{{Martin}}, \binits{P.-E.}},
\bauthor{\bsnm{{Sephton}}, \binits{M.A.}},
\bauthor{\bsnm{{Watson}}, \binits{J.S.}},
\bauthor{\bsnm{{Salge}}, \binits{T.}},
\bauthor{\bsnm{{Shirley}}, \binits{K.A.}},
\bauthor{\bsnm{{Curtis}}, \binits{R.J.}},
\bauthor{\bsnm{{Warren}}, \binits{T.J.}},
\bauthor{\bsnm{{Bowles}}, \binits{N.E.}},
\bauthor{\bsnm{{Stuart}}, \binits{F.M.}},
\bauthor{\bsnm{{Di Nicola}}, \binits{L.}},
\bauthor{\bsnm{{Gy{\"o}re}}, \binits{D.}},
\bauthor{\bsnm{{Boyce}}, \binits{A.J.}},
\bauthor{\bsnm{{Shaw}}, \binits{K.M.M.}},
\bauthor{\bsnm{{Elliott}}, \binits{T.}},
\bauthor{\bsnm{{Steele}}, \binits{R.C.J.}},
\bauthor{\bsnm{{Povinec}}, \binits{P.}},
\bauthor{\bsnm{{Laubenstein}}, \binits{M.}},
\bauthor{\bsnm{{Sanderson}}, \binits{D.}},
\bauthor{\bsnm{{Cresswell}}, \binits{A.}},
\bauthor{\bsnm{{Jull}}, \binits{A.J.T.}},
\bauthor{\bsnm{{S{\'y}kora}}, \binits{I.}},
\bauthor{\bsnm{{Sridhar}}, \binits{S.}},
\bauthor{\bsnm{{Harrison}}, \binits{R.J.}},
\bauthor{\bsnm{{Willcocks}}, \binits{F.M.}},
\bauthor{\bsnm{{Harrison}}, \binits{C.S.}},
\bauthor{\bsnm{{Hallatt}}, \binits{D.}},
\bauthor{\bsnm{{Wozniakiewicz}}, \binits{P.J.}},
\bauthor{\bsnm{{Burchell}}, \binits{M.J.}},
\bauthor{\bsnm{{Alesbrook}}, \binits{L.S.}},
\bauthor{\bsnm{{Dignam}}, \binits{A.}},
\bauthor{\bsnm{{Almeida}}, \binits{N.V.}},
\bauthor{\bsnm{{Smith}}, \binits{C.L.}},
\bauthor{\bsnm{{Clark}}, \binits{B.}},
\bauthor{\bsnm{{Humphreys-Williams}}, \binits{E.R.}},
\bauthor{\bsnm{{Schofield}}, \binits{P.F.}},
\bauthor{\bsnm{{Cornwell}}, \binits{L.T.}},
\bauthor{\bsnm{{Spathis}}, \binits{V.}},
\bauthor{\bsnm{{Morgan}}, \binits{G.H.}},
\bauthor{\bsnm{{Perkins}}, \binits{M.J.}},
\bauthor{\bsnm{{Kacerek}}, \binits{R.}},
\bauthor{\bsnm{{Campbell-Burns}}, \binits{P.}},
\bauthor{\bsnm{{Colas}}, \binits{F.}},
\bauthor{\bsnm{{Zanda}}, \binits{B.}},
\bauthor{\bsnm{{Vernazza}}, \binits{P.}},
\bauthor{\bsnm{{Bouley}}, \binits{S.}},
\bauthor{\bsnm{{Jeanne}}, \binits{S.}},
\bauthor{\bsnm{{Hankey}}, \binits{M.}},
\bauthor{\bsnm{{Collins}}, \binits{G.S.}},
\bauthor{\bsnm{{Young}}, \binits{J.S.}},
\bauthor{\bsnm{{Shaw}}, \binits{C.}},
\bauthor{\bsnm{{Horak}}, \binits{J.}},
\bauthor{\bsnm{{Jones}}, \binits{D.}},
\bauthor{\bsnm{{James}}, \binits{N.}},
\bauthor{\bsnm{{Bosley}}, \binits{S.}},
\bauthor{\bsnm{{Shuttleworth}}, \binits{A.}},
\bauthor{\bsnm{{Dickinson}}, \binits{P.}},
\bauthor{\bsnm{{McMullan}}, \binits{I.}},
\bauthor{\bsnm{{Robson}}, \binits{D.}},
\bauthor{\bsnm{{Smedley}}, \binits{A.R.D.}},
\bauthor{\bsnm{{Stanley}}, \binits{B.}},
\bauthor{\bsnm{{Bassom}}, \binits{R.}},
\bauthor{\bsnm{{McIntyre}}, \binits{M.}},
\bauthor{\bsnm{{Suttle}}, \binits{A.A.}},
\bauthor{\bsnm{{Fleet}}, \binits{R.}},
\bauthor{\bsnm{{Bastiaens}}, \binits{L.}},
\bauthor{\bsnm{{Ih{\'a}sz}}, \binits{M.B.}},
\bauthor{\bsnm{{McMullan}}, \binits{S.}},
\bauthor{\bsnm{{Boazman}}, \binits{S.J.}},
\bauthor{\bsnm{{Dickeson}}, \binits{Z.I.}},
\bauthor{\bsnm{{Grindrod}}, \binits{P.M.}},
\bauthor{\bsnm{{Pickersgill}}, \binits{A.E.}},
\bauthor{\bsnm{{Weir}}, \binits{C.J.}},
\bauthor{\bsnm{{Suttle}}, \binits{F.M.}},
\bauthor{\bsnm{{Farrelly}}, \binits{S.}},
\bauthor{\bsnm{{Spencer}}, \binits{I.}},
\bauthor{\bsnm{{Naqvi}}, \binits{S.}},
\bauthor{\bsnm{{Mayne}}, \binits{B.}},
\bauthor{\bsnm{{Skilton}}, \binits{D.}},
\bauthor{\bsnm{{Kirk}}, \binits{D.}},
\bauthor{\bsnm{{Mounsey}}, \binits{A.}},
\bauthor{\bsnm{{Mounsey}}, \binits{S.E.}},
\bauthor{\bsnm{{Mounsey}}, \binits{S.}},
\bauthor{\bsnm{{Godfrey}}, \binits{P.}},
\bauthor{\bsnm{{Bond}}, \binits{L.}},
\bauthor{\bsnm{{Bond}}, \binits{V.}},
\bauthor{\bsnm{{Wilcock}}, \binits{C.}},
\bauthor{\bsnm{{Wilcock}}, \binits{H.}},
\bauthor{\bsnm{{Wilcock}}, \binits{R.}}:
\batitle{{The Winchcombe meteorite, a unique and pristine witness from the
  outer solar system}}.
\bjtitle{Science Advances}
\bvolume{8}(\bissue{46}),
\bfpage{3925}
(\byear{2022})
\doiurl{10.1126/sciadv.abq3925}
\end{barticle}
\endbibitem

\bibitem[\protect\citeauthoryear{{O'Brien} et~al.}{2022}]{OBrien+22}
\begin{barticle}
\bauthor{\bsnm{{O'Brien}}, \binits{{\'A}.C.}},
\bauthor{\bsnm{{Pickersgill}}, \binits{A.}},
\bauthor{\bsnm{{Daly}}, \binits{L.}},
\bauthor{\bsnm{{Jenkins}}, \binits{L.}},
\bauthor{\bsnm{{Floyd}}, \binits{C.}},
\bauthor{\bsnm{{Martin}}, \binits{P.-E.}},
\bauthor{\bsnm{{Hallis}}, \binits{L.J.}},
\bauthor{\bsnm{{King}}, \binits{A.}},
\bauthor{\bsnm{{Lee}}, \binits{M.}}:
\batitle{{The Winchcombe Meteorite: one year on}}.
\bjtitle{Astronomy and Geophysics}
\bvolume{63}(\bissue{1}),
\bfpage{1}--\blpage{21123}
(\byear{2022})
\doiurl{10.1093/astrogeo/atac009}
\end{barticle}
\endbibitem

\bibitem[\protect\citeauthoryear{Koepnick et~al.}{2019}]{Koepnick2019}
\begin{barticle}
\bauthor{\bsnm{Koepnick}, \binits{B.}},
\bauthor{\bsnm{Flatten}, \binits{J.}},
\bauthor{\bsnm{Husain}, \binits{T.}},
\bauthor{\bsnm{Ford}, \binits{A.}},
\bauthor{\bsnm{Silva}, \binits{D.-A.}},
\bauthor{\bsnm{Bick}, \binits{M.J.}},
\bauthor{\bsnm{Bauer}, \binits{A.}},
\bauthor{\bsnm{Liu}, \binits{G.}},
\bauthor{\bsnm{Ishida}, \binits{Y.}},
\bauthor{\bsnm{Boykov}, \binits{A.}},
\bauthor{\bsnm{Estep}, \binits{R.D.}},
\bauthor{\bsnm{Kleinfelter}, \binits{S.}},
\bauthor{\bsnm{N{\o}rg{\aa}rd-Solano}, \binits{T.}},
\bauthor{\bsnm{Wei}, \binits{L.}},
\bauthor{\bsnm{Players}, \binits{F.}},
\bauthor{\bsnm{Montelione}, \binits{G.T.}},
\bauthor{\bsnm{DiMaio}, \binits{F.}},
\bauthor{\bsnm{Popovi{\'{c}}}, \binits{Z.}},
\bauthor{\bsnm{Khatib}, \binits{F.}},
\bauthor{\bsnm{Cooper}, \binits{S.}},
\bauthor{\bsnm{Baker}, \binits{D.}}:
\batitle{De novo protein design by citizen scientists}.
\bjtitle{Nature}
\bvolume{570}(\bissue{7761}),
\bfpage{390}--\blpage{394}
(\byear{2019})
\doiurl{10.1038/s41586-019-1274-4}
\end{barticle}
\endbibitem

\bibitem[\protect\citeauthoryear{Fraenkel}{1993}]{Fraenkel1993}
\begin{barticle}
\bauthor{\bsnm{Fraenkel}, \binits{A.S.}}:
\batitle{Complexity of protein folding}.
\bjtitle{Bulletin of Mathematical Biology}
\bvolume{55}(\bissue{6}),
\bfpage{1199}--\blpage{1210}
(\byear{1993})
\doiurl{10.1007/bf02460704}
\end{barticle}
\endbibitem

\end{thebibliography}

\end{document}